\pdfoutput=1
\documentclass[aps,prd,amsmath,floats,floatfix, twocolumn,
superscriptaddress,nofootinbib,showpacs,longbibliography]{revtex4-1}

\usepackage[T1]{fontenc}
\usepackage[utf8]{inputenc}
\usepackage{lmodern}

\usepackage{verbatim}

\usepackage[dvipsnames, usenames]{xcolor}
\definecolor{linkcolor}{rgb}{0.0,0.3,0.5}
\usepackage[hypertexnames=false, unicode, colorlinks=true, linkcolor=linkcolor,
citecolor=linkcolor, filecolor=linkcolor,urlcolor=linkcolor,
pdfusetitle]{hyperref}

\usepackage[all]{hypcap}
\usepackage{graphicx}
\usepackage{xspace}
\usepackage{braket}
\usepackage{amssymb}
\usepackage[normalem]{ulem} 
\usepackage{bm} 

\usepackage{microtype}

\usepackage[english]{babel}
\usepackage{blindtext}
\usepackage{mathrsfs}
\graphicspath{%
  {figs/}%
}

\DeclareMathAlphabet{\mathpzc}{OT1}{pzc}{m}{it}

\newcommand{\h}{\mathpzc{h}}
\newcommand{\hlm}{\mathpzc{h}_{\ell m}}
\newcommand{\chieff}{\chi_{\mathrm{eff}}}

\newcommand{\bchi}{\bm{\chi}}

\newcommand{\fLow}{f_{\mathrm{low}}}
\newcommand{\mSrc}[1]{m^{\mathrm{src}}_{#1}}
\newcommand{\MSrc}{M^{\mathrm{src}}}
\newcommand{\MchirpSrc}{\mathcal{M}^{\mathrm{src}}}

\newcommand{\NRSurNew}{\texttt{NRHybSur2dq15}\xspace}
\newcommand{\NRSurNewOnlyNR}{\texttt{NRSur2dq15}\xspace}
\newcommand{\NRSurOld}{\texttt{NRHybSur3dq8}\xspace}
\newcommand{\EOBnoHM}{\texttt{SEOBNRv4}\xspace}
\newcommand{\EOBHM}{\texttt{SEOBNRv4HM}\xspace}
\newcommand{\EOBPHM}{\texttt{SEOBNRv4PHM}\xspace}
\newcommand{\IMRTHM}{\texttt{IMRPhenomTHM}\xspace}

\newcommand{\IMRTPHM}{\texttt{IMRPhenomTPHM}\xspace}
\newcommand{\IMRPHMOld}{\texttt{IMRPhenomPv3PHM}\xspace}

\begin{document}

\title{Targeted large mass ratio numerical relativity surrogate waveform model for GW190814}

\newcommand\Caltech{\affiliation{TAPIR 350-17, California Institute of
Technology, 1200 E California Boulevard, Pasadena, CA 91125, USA}}
\newcommand{\AEI}{\affiliation{Max Planck Institute for Gravitational Physics
(Albert Einstein Institute), D-14476 Potsdam, Germany}}
\newcommand{\Cornell}{\affiliation{Cornell Center for Astrophysics and Planetary Science,
Cornell University, Ithaca, New York 14853, USA}}
\newcommand\CornellPhys{\affiliation{Department of Physics, Cornell
    University, Ithaca, New York 14853, USA}}
\newcommand{\CITA}{\affiliation{Canadian Institute for Theoretical
    Astrophysics, 60 St.~George Street, University of Toronto,
    Toronto, ON M5S 3H8, Canada}} %
\newcommand{\GWPAC}{\affiliation{Gravitational Wave Physics and
    Astronomy Center, California State University Fullerton,
    Fullerton, California 92834, USA}} %
\newcommand{\UMassD}{\affiliation{Department of Mathematics,
    Center for Scientific Computing and Visualization Research,
    University of Massachusetts, Dartmouth, MA 02747, USA}}
\newcommand\olemiss{\affiliation{Department of Physics and Astronomy,
The University of Mississippi, University, MS 38677, USA}}
\newcommand{\bham}{\affiliation{School of Physics and Astronomy and Institute
for Gravitational Wave Astronomy, University of Birmingham, Birmingham, B15
2TT, UK}}
\newcommand\MIT{\affiliation{LIGO Laboratory, Massachusetts Institute of
Technology, Cambridge, Massachusetts 02139, USA}}
\newcommand{\MKI}{\affiliation{Department of Physics and Kavli Institute for Astrophysics and Space Research, Massachusetts Institute of Technology, 77 Massachusetts Ave, Cambridge, MA 02139, USA}}

\author{Jooheon Yoo}
\email{jy884@cornell.edu}
\Cornell

\author{Vijay Varma}
\email{vijay.varma@aei.mpg.de}
\thanks{Marie Curie Fellow}
\AEI
\CornellPhys
\Cornell

\author{Matthew Giesler}
\Cornell

\author{Mark A. Scheel}
\Caltech

\author{Carl-Johan Haster}
\MIT
\MKI

\author{Harald P. Pfeiffer}
\AEI

\author{Lawrence E. Kidder}
\Cornell

\author{Michael Boyle}
\Cornell

\hypersetup{pdfauthor={Yoo et al.}}

\date{\today}

\begin{abstract}
Gravitational wave observations of large mass ratio compact binary mergers like
GW190814 highlight the need for reliable, high-accuracy waveform templates for
such systems. We present \NRSurNew, a new surrogate model trained on hybridized
numerical relativity (NR) waveforms with mass ratios $q\leq15$, and aligned
spins $|\chi_{1z}|\leq0.5$ and $\chi_{2z}=0$. We target the parameter space of
GW190814-like events as large mass ratio NR simulations are very expensive. The
model includes the (2,2), (2,1), (3,3), (4,4), and (5,5) spin-weighted
spherical harmonic modes, and spans the entire LIGO-Virgo bandwidth (with $\fLow=20$
Hz) for total masses $M \gtrsim 9.5 \, M_{\odot}$. \NRSurNew accurately
reproduces the hybrid waveforms, with mismatches below $\sim 2 \times 10^{-3}$
for total masses $10 \, M_{\odot} \leq M \leq 300 \, M_{\odot}$. This is at
least an order of magnitude improvement over existing semi-analytical models
for GW190814-like systems. Finally, we reanalyze GW190814 with the new model
and obtain source parameter constraints consistent with previous work.
\end{abstract}

\maketitle


\section{Introduction}
\label{sec:intro}

The LIGO~\cite{TheLIGOScientific:2014jea} and Virgo~\cite{TheVirgo:2014hva}
detectors have observed a total of $90$ gravitational wave (GW) signals to
date~\cite{LIGOScientific:2018mvr, LIGOScientific:2021usb, LIGOScientific:2021djp}, including the landmark observations of the
first binary black hole (BH)~\cite{Abbott:2016blz}, binary neutron star
(NS)~\cite{TheLIGOScientific:2017qsa}, and BH-NS
binaries~\cite{LIGOScientific:2021qlt}. Among these observations,
GW190814~\cite{Abbott:2020khf} is unique due to its uncertain nature: a merger
of a $\sim23 \, M_{\odot}$ BH and a $\sim2.6 \, M_{\odot}$ companion that is
either the heaviest NS or the lightest BH ever discovered~\cite{Abbott:2020khf}
in a compact binary system.\footnote{A similar event, GW200210\_092254, a
merger of a $24.1 \, M_{\odot}$ BH and a $2.81 \, M_{\odot}$ compact object was
identified in Ref.~\cite{LIGOScientific:2021djp}.  However, this event is a
marginal GW candidate, with a probability of astrophysical origin
$p_{\text{astro}} \sim 0.54$~\cite{LIGOScientific:2021djp}. Therefore, we limit
our analysis to GW190814.} In addition to the intrigue about its astrophysical
origin~\cite{Godzieba:2020tjn, Dexheimer:2020rlp, Clesse:2020ghq, Tews:2020ylw,
Tsokaros:2020hli, Fattoyev:2020cws, Zhang:2020zsc, Tan:2020ics, Nathanail:2021tay}, this event
also poses new challenges for waveform models due to the highly unequal masses
of the binary components.

Numerical relativity (NR) is the only available method for solving Einstein's
equations near the merger of two compact objects, and has played a central role
in GW astronomy~\cite{Pretorius:2005gq, Campanelli:2005dd, Baker:2005vv,
Boyle:2019kee}. Unfortunately, NR simulations are prohibitively expensive
for direct GW data analysis applications, as each simulation can take up to a
few months on a supercomputer. The need for a faster alternative to NR has led
to the development of several semi-analytical waveform
models~\cite{Ossokine:2020kjp, Khan:2019kot, Matas:2020wab, Thompson:2020nei,
    Cotesta:2018fcv, Estelles:2021gvs, Estelles:2020twz, Pratten:2020ceb,
Garcia-Quiros:2020qpx, Akcay:2020qrj, Nagar:2020pcj} that rely on some
physically motivated assumptions for the underlying phenomenology, and
calibrate the remaining free parameters to NR simulations.  As a result, these
models are fast enough for GW data analysis, but are typically not as accurate
as the NR simulations~\cite{Varma:2021csh, Varma:2018mmi, Varma:2019csw}.

On the other hand, NR surrogate models~\cite{Field:2013cfa, Blackman:2017pcm,
Varma:2018mmi, Varma:2019csw} take a data-driven approach by training the model
directly on NR simulations, without the need for added assumptions. These
models have been shown to reproduce NR simulations without a significant loss
of accuracy while also being fast enough for GW data
analysis~\cite{Varma:2018mmi, Varma:2019csw}. The main limitation for surrogate
models, however, is that their applicability is restricted to the regions where
sufficient NR simulations are available. In particular, NR simulations become
expensive as one approaches large mass ratios $q=m_1/m_2$ and/or large spin
magnitudes $\chi_{1,2}$~\cite{Boyle:2019kee, Lousto:2020tnb}, where $m_{1}$
($m_2$) represents the mass of the heavier (lighter) BH, so that $q\geq1$, and
$\bchi_{1,2}$ represent the corresponding dimensionless spins, with magnitudes
$\chi_{1,2}\leq 1$.  Therefore, previous NR surrogate models have only been
trained on simulations with $q\leq8$ and $\chi_{1,2} \leq
0.8$~\cite{Varma:2018mmi}. These models are not suitable for high-mass ratio
systems like GW190814 ($q \sim 8.96^{+0.75}_{-0.62}$ at 90\%
credibility~\cite{Abbott:2020khf}).

Similarly, the calibration NR data for the semi-analytical
models~\cite{Ossokine:2020kjp, Khan:2019kot, Matas:2020wab, Thompson:2020nei}
used in the GW190814 discovery paper~\cite{Abbott:2020khf} are also very sparse
at mass ratios $q \gtrsim 8$. Fortunately, most of the events observed by
LIGO-Virgo fall at more moderate mass ratios $q\lesssim
5$~\cite{LIGOScientific:2021djp}, with a preference for $q\sim 1$
~\cite{LIGOScientific:2021psn}, where current semi-analytical models are well
calibrated. In contrast, the large mass ratio of GW190814 poses new challenges
for waveform modeling, and it is important to understand the impact of modeling
error on the source parameter estimation of this event.

For example, at large $q$, subdominant modes of radiation beyond the quadrupole
mode can play an important role. The complex waveform $\h = h_{+} - i
h_{\times}$ can be decomposed into a sum of spin-weighted spherical harmonic
modes $\hlm$:
\begin{align}
\h(t,\iota,\varphi_0) = \sum^\infty_{l=2} \sum^{\ell}_{m=-\ell} \hlm(t)
                        \ _{-2}Y_{\ell m}(\iota,\varphi_0),
\label{eq:hlm}
\end{align}
where $h_{+}$ ($h_{\times}$) represents the plus (cross) GW polarization,
$_{-2}Y_{\ell m}$ are the spin $=-2$ weighted spherical harmonics, and
$(\iota,\varphi_0)$ represent the direction to the observer in the source
frame.\footnote{The source frame is defined as follows: the $z$-axis points
along the orbital angular momentum $\bm{L}$ of the binary, the $x$-axis points
along the line of separation from the lighter BH to the heavier BH, and the
$y$-axis completes the triad. Therefore, $\iota$ denotes the inclination angle
between $\bm{L}$ and line-of-sight to the observer.} The $\ell=|m|= 2$ terms
typically dominate the sum in Eq.~(\ref{eq:hlm}), and are referred to as the
quadrupole modes. However, as one approaches large $q$ the subdominant modes
(also referred to as nonquadrupole or higher modes) become increasingly
important for estimating the binary source properties~\cite{Varma:2016dnf,
Varma:2014jxa, Capano:2013raa, Shaik:2019dym, Islam:2021zee}. Therefore, it is
important for waveform models to accurately capture the effect of the
subdominant modes on the observed signal. Along with developing a new surrogate
model, one of the goals of this work is to assess whether current
semi-analytical models, and in particular their subdominant modes, are accurate
enough for events like GW190814.

\subsection{The \NRSurNew model}

In this work, we build a GW190814-targeted surrogate model that is based on NR
simulations with mass ratios up to $q=15$. Due to the computational cost of NR
simulations with large mass ratios and/or spins~\cite{Boyle:2019kee}, we
restrict the model to spins (anti-) aligned along the direction of the orbital
angular momentum $\bm{L}$, with $\chi_{1z} \in [-0.5, 0.5]$,
$\chi_{1x}=\chi_{1y}=0$, and $\bchi_{2}=0$.
We ignore the spin of the secondary BH for simplicity, as its effect is
expected to be suppressed for large $q$ systems like GW190814, at least at
current signal to noise ratio (SNR). For example, Ref.~\cite{Abbott:2020khf}
found that the secondary spin of GW190814 was unconstrained. This assumption
may need to be relaxed for louder signals that are expected in the future with
detector improvements.

Above, the $z$-direction is taken to be along $\bm{L}$, whose direction is
constant for aligned-spin systems. In addition to the dominant ($(\ell, m) =
(2,2)$) mode, the model accurately captures effects of the following
subdominant modes: (2,1), (3,3), (4,4) and (5,5). Note that the $m<0$ modes
carry the same information as $m>0$ modes for aligned-spin binaries, and do not
need to be modeled separately.

To train the model, we perform 20 new NR simulations in the range $8<q\leq15$,
using the Spectral Einstein Code (SpEC)~\cite{SpECwebsite, Boyle:2019kee}
developed by the SXS~\cite{SXSWebsite} collaboration.  Due to computational
limitations, these simulations only include about 30 orbits before the merger;
therefore, they do not cover the full LIGO-Virgo frequency band for stellar
mass binaries. More precisely, for total masses $M=m_1+m_2 \lesssim
70.0\,M_{\odot}$, the initial frequency of the $(2,2)$ mode of these waveforms
falls within the LIGO-Virgo band, taken to begin at $\fLow=20$ Hz. We extend the
validity of the model to lower masses by smoothly
transitioning~\cite{Varma:2018mmi} to the effective-one-body (EOB) model
\EOBHM~\cite{Cotesta:2018fcv} for the early inspiral. These NR-EOB
\emph{hybrid} waveforms are augmented with $31$ waveforms in the $q\leq8$
region, generated using the \NRSurOld~\cite{Varma:2018mmi} surrogate model,
which is already hybridized.  The new model, \NRSurNew is trained on these $51$
hybrid waveforms, and all modes of this model are valid for full LIGO-Virgo band
(with $\fLow=20$ Hz) for $M \gtrsim 9.5 \, M_{\odot}$.

For simplicity, \NRSurNew ignores two physical features that can be relevant
for GW190814: precession and tidal deformability of the secondary object.
Precession occurs when the component objects have spins that are tilted with
respect to $\bm{L}$. In such binaries, the spins interact with $\bm{L}$ (as
well as with each other), causing the orbital plane to
precess~\cite{Apostolatos:1994pre}. The effective precession parameter
$\chi_{\text{p}}$~\cite{Schmidt:2014iyl} for GW190814 was constrained to
$\chi_{\text{p}} \lesssim 0.07$ at 90\% credibility by
Ref.~\cite{Abbott:2020khf}. However, including precession in the waveform model
was found to improve the component mass constraints~\cite{Abbott:2020khf}.
Therefore, while neglecting precession is a reasonable assumption, this can
limit the applicability of our results. Precessing NR surrogates can require
$\gtrsim 1000$ NR simulations~\cite{Blackman:2017dfb, Blackman:2017pcm,
Varma:2019csw}, which is not currently feasible for large mass
ratios~\cite{Boyle:2019kee}. Nevertheless, we can still compare the performance
of \NRSurNew against other nonprecessing models.

Next, the tidal deformations of NSs within a compact binary can alter the
orbital dynamics, imprinting a signature on the GW
signal~\cite{Flanagan:2007ix}. Assuming the secondary object of GW190814 is a
NS, this effect, parameterized by the effective tidal
deformability~\cite{Flanagan:2007ix} scales as $\tilde{\Lambda} \propto 1/q^4$
(see e.g.~Eq.~(1) of Ref.~\cite{TheLIGOScientific:2017qsa}), and can be safely
ignored for GW190814~\cite{Abbott:2020khf}. For large $q$ binaries like
GW190814, the NS simply plunges into the BH before tidal deformation or
disruption can occur~\cite{Foucart:2013psa}. As a result, GW190814 shows no
evidence of measurable tidal effects in the signal, and no electromagnetic
counterpart to the GWs has been identified~\cite{Abbott:2020khf}. This
justifies our choice to ignore the effects of tidal deformation in \NRSurNew.

To summarize, \NRSurNew is valid for mass ratios $q\leq15$, spins $\chi_{1z}
\in [-0.5, 0.5]$ and $\chi_{1x}=\chi_{1y}=\bchi_{2}=0$, total masses $M \gtrsim
9.5 \, M_{\odot}$ (for $\fLow=20$ Hz), and zero tidal deformability. The name
of the model is derived from the fact that it is based on NR hybrid waveforms,
spans the 2-dimensional parameter space of $(q, \chi_{1z})$, and extends to
$q=15$.

The rest of the paper is organized as follows. In Sec.~\ref{sec:method}, we
describe the construction of \NRSurNew. In Sec.~\ref{sec:surrogate_errors}, we
evaluate the accuracy of the model by computing mismatches against NR-EOB
hybrid waveforms. We demonstrate that \NRSurNew is more accurate than existing
semi-analytical models by at least an order of magnitude, with mismatches
$\lesssim 2 \times 10^{-3}$ throughout its parameter space. In
Sec.~\ref{sec:reanalyzing_GW190814}, we reanalyze GW190814 using \NRSurNew and
find that our constraints on the binary properties are consistent with those
reported in Ref.~\cite{Abbott:2020khf}.  We end with some concluding remarks in
Sec.~\ref{sec:conc}.  Throughout this paper, we denote redshifted the detector
frame masses as $m_1$, $m_2$, and $M=m_1+m_2$. When referring to the source
frame masses, we denote them explicitly as $\mSrc{1}$, $\mSrc{2}$, and $\MSrc$.
These are related by factors of $1+z$, where $z$ is the cosmological redshift;
for example, $M=(1+z) \, \MSrc$.

\section{Methods}
\label{sec:method}

In this section we describe the steps involved in building the new model
\NRSurNew, including the generation of the required NR and hybrid waveforms,
and the surrogate model construction.

\subsection{Training set generation}
\label{sec:training_set}

In order to build the surrogate model, we need a \emph{training set} of hybrid
waveforms and their associated binary parameters. The parameter space of
interest for us is the 2D region $q \in [1, 15]$ and $\chi_{1z} \in [-0.5,
0.5]$, with fixed $\chi_{1x}=\chi_{1y}=0$, and $\bchi_{2}=0$.  The total mass
scales out for binary BHs and does not need to be modeled separately. The NR
simulations necessary for generating hybrid waveforms are expensive, especially
as one approaches large $q$~\cite{Boyle:2019kee}.  Therefore, one would ideally
like to use the fewest possible hybrid waveforms to build a surrogate model of
given a target accuracy. However, we do not know a priori how big the training
set should be or how these points should be distributed in the parameter space.
In order to determine a suitable training set, we first build a surrogate model
for post-Newtonian (PN) waveforms.

\subsubsection{PN surrogate and new NR simulations}
\label{sec:pn_sur}

We use the GWFrames package~\cite{GWFrames} to generate PN waveforms. For the
orbital phase, we use the TaylorT4~\cite{Boyle:2007ft} approximant, and include
nonspinning terms up to 4 PN order~\cite{Blanchet:2004ek, Blanchet:2004ek,
Jaranowski:2013lca, Bini:2013zaa, Bini:2013rfa} and spin terms up to 2.5 PN
order~\cite{Kidder:1995zr, Will:1996zj, Bohe:2012mr}. For the amplitudes, we
include terms up to 3.5 PN order~\cite{Blanchet:2008je, Faye:2012we,
Faye:2014fra}. For the PN surrogate, we restrict the length of the waveforms to
be $5000 \, M$, terminating at the orbital frequency of the Schwarzschild
innermost-stable-circular-orbit (ISCO): $\omega_{\mathrm{orb}} = 6^{-3/2} \,
\mathrm{rad}/M$. In addition, we only use the $(2,2)$ mode for simplicity.
Despite the restrictions in length, mode-content, and the missing
merger-ringdown section in the PN waveforms, we find that this approach
provides a good initial training set for constructing hybrid NR-EOB
surrogates~\cite{Varma:2018mmi}. Above, the orbital frequency is defined as:
\begin{equation}
    \omega_{\mathrm{orb}} = \frac{d\phi_{\mathrm{orb}}}{dt},
\label{eq:omega_orb}
\end{equation}
where $\phi_{\mathrm{orb}}$ is the orbital phase obtained from the (2,2) mode
(see Eq.~(\ref{eq:orb_phase})).

\begin{figure}[tbh]
\centering
\includegraphics[width=0.45\textwidth]{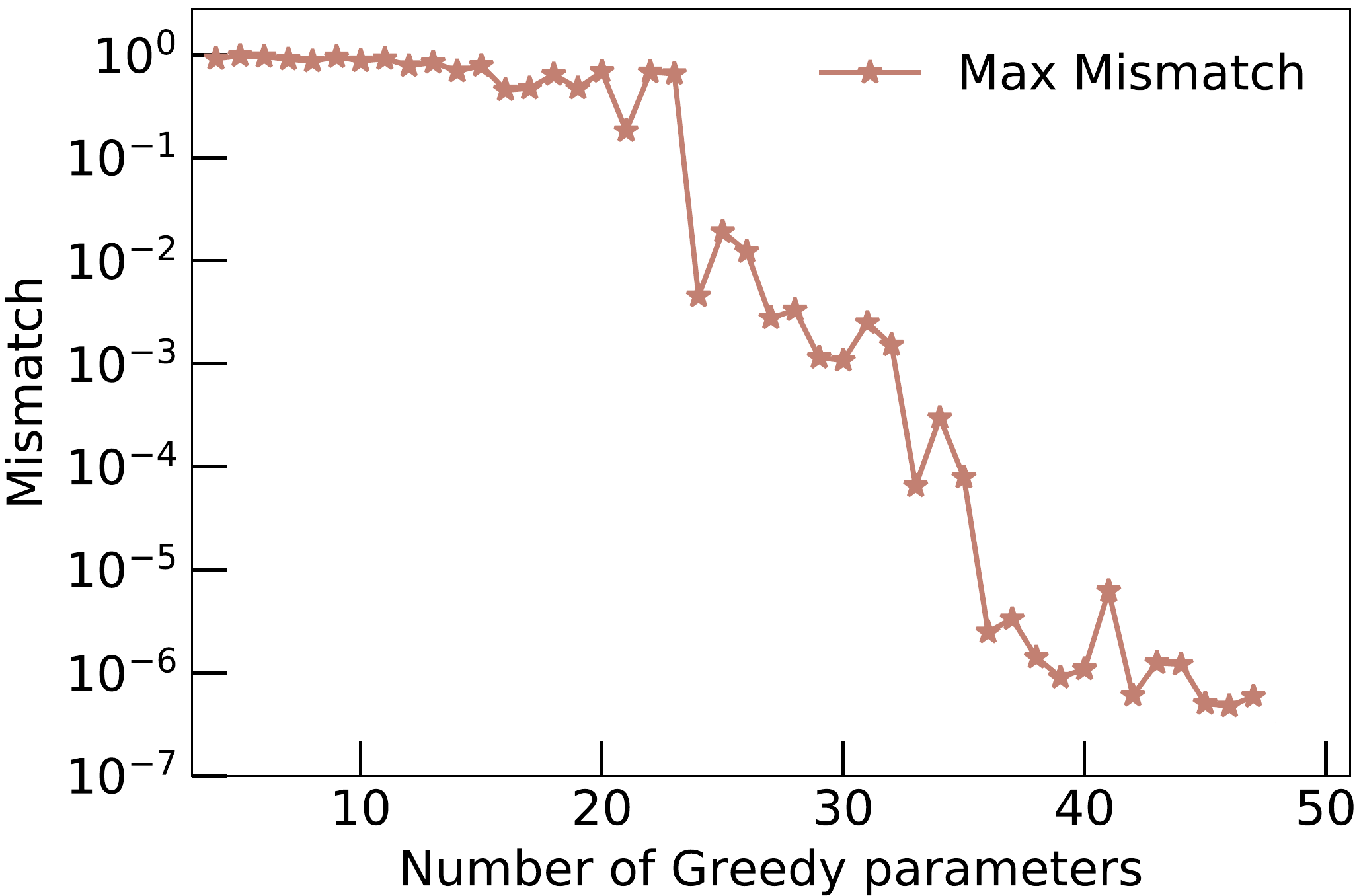}
\caption{
Largest mismatch of the PN surrogate (over the entire validation set) as a
function of number of greedy parameters used for training. The PN surrogate is
seen to converge to the validation waveforms as the size of the training set
increases.
}
\label{fig:pn_greedy}
\end{figure}

We initialize the training set for the PN surrogate with just the corner cases
of the parameter space. For our 2D model, these consist of the four points:
$(q, \, \chi_{1z}) = (1, \, \pm 0.5)$ and $(15, \, \pm 0.5)$.  We augment the
training set in an iterative \emph{greedy} manner: At each iteration, we build
a PN surrogate with the current training set, following the same methods as we
use for the hybrid surrogate (see Sec.~\ref{sec:surrogate_building}). Then, we
test this surrogate against a larger ($\sim 10$ times) \emph{validation set},
generated by randomly sampling the parameter space at each
iteration.\footnote{
The boundary parameters are expected to be more important than those in the
bulk; therefore, for 30\% of the points in the validation set, we sample only
from the boundary, which corresponds to the edges of a square in the 2D case.}
We select the parameter in the validation set that has
the largest error (computed using Eq.~(\ref{eq:mismatch_td})) and add it to the
training set for the next iteration. We repeat this procedure until the largest
validation error falls below a certain threshold.

In order to estimate the error between two complex waveforms $\h_1$ and $\h_2$,
we use the time-domain inner product,
\begin{align}
    & \braket{\h_1,\h_2} = \bigg| \int_{t_{\mathrm{min}}}^{t_{\mathrm{max}}} \h_1(t) \h_2^*(t) dt \bigg| ,
\end{align}
to compute the mismatch,
\begin{align}
    & \mathcal{MM} = 1 - \frac{\braket{\h_1,\h_2}}{\sqrt{\braket{\h_1,\h_1}\braket{\h_2,\h_2}}}
\label{eq:mismatch_td}
\end{align}
When computing mismatches for the PN surrogate, we assume a flat noise curve,
and do not optimize over time and phase shifts.

Figure~\ref{fig:pn_greedy} shows the maximum validation error at each iteration
against the size of the training set. We stop this procedure when the training
set size reaches 47, as the mismatch settles below $10^{-6}$ at this point.
Among these, 31 cases lie in the region $q\leq8$, while 16 lie in the region
$8<q\leq15$. Rather than perform new NR simulations for the $q\leq8$ cases, we
generate waveforms using the existing \NRSurOld model~\cite{Varma:2018mmi}.
This model was trained on NR-EOB/PN hybrid waveforms with mass ratios $q\leq8$
and spins $\chi_{1z,2z}\in [-0.8, 0.8]$, and was shown to reproduce the hybrid
waveforms without a significant loss of accuracy~\cite{Varma:2018mmi}.

For the cases with $q>8$, we perform new NR simulations using
SpEC~\cite{SpECwebsite, Boyle:2019kee}. These NR waveforms include $\sim 5000M$
of evolution before the merger and are hybridized using
\EOBHM~\cite{Cotesta:2018fcv} waveforms to include the early inspiral (see
Sec.~\ref{sec:hybridization}). However, of the 16 cases with $q>8$, only 15
simulations were successfully completed.\footnote{The reason for failure is
large constraint violation as the binary approaches merger. We believe a better
domain decomposition may be needed for this simulation, which we plan to
explore in the future.} This leaves us with a total of 46 training waveforms
(15 NR-EOB hybrid waveforms and 31 \NRSurOld waveforms).

From an initial attempt to build a hybrid surrogate with these 46 waveforms, we
found that the model performs poorly for low masses $\lesssim 50 M_{\odot}$,
with mismatches reaching $\sim 10^{-2}$, but performs very well for higher
masses, with mismatches $\sim 10^{-3}$. In other words, the late inspiral and
merger-ringdown stages were accurately captured, but the early inspiral was
not. This suggested that more hybrid waveforms were required. To estimate where
in parameter space to place new hybrid waveforms, we first constructed a trial
NR-only surrogate using the above training set of 46 waveforms, but restricted
to the last $5000M$ before merger; we will refer to this model as
\NRSurNewOnlyNR. Next, we hybridized waveforms (see
Sec.~\ref{sec:hybridization}) obtained from \NRSurNewOnlyNR to generate new
training points in the $q>8$ region. This bootstrap method allowed us to create
as many hybrid waveforms as necessary in the $q>8$ region without performing
new NR simulations. After some trial and error, we found that placing five new
hybrid waveforms at $q=14$ (uniformly distributed in $\chi_{1z}\in[-0.5, 0.5]$)
resolved the problem at low masses.

\begin{figure}[tbh]
\centering
\includegraphics[width=8.4cm]{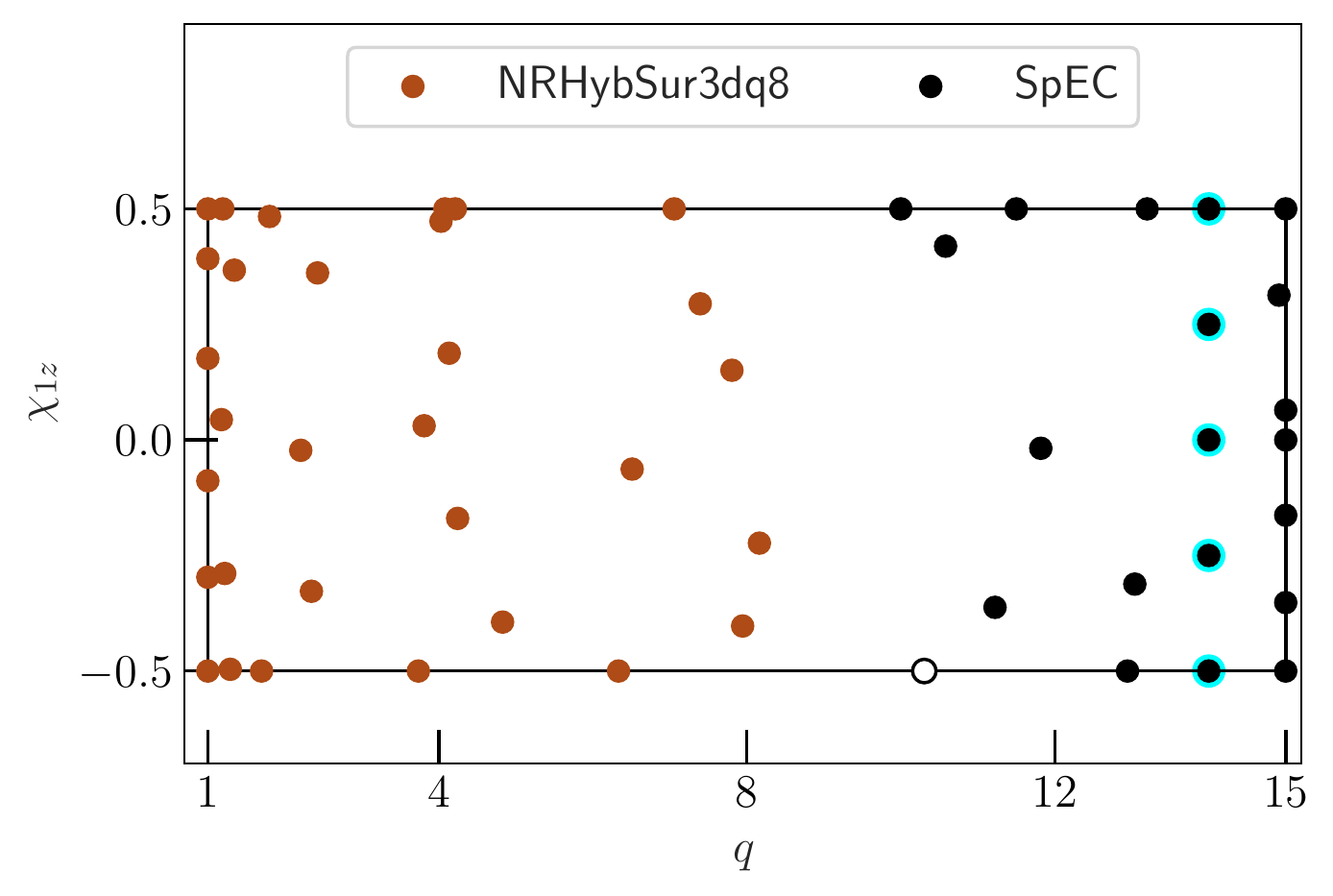}
\caption{
Training set parameters used in building the surrogate model \NRSurNew. The red
markers correspond to cases with $q\leq8$, for which \NRSurOld is used to
generate training waveforms. The black markers represent the new NR waveforms
performed for this work, while the empty marker shows the failed NR simulation.
The distribution of the 47 parameters from Fig.~\ref{fig:pn_greedy} can be seen
by ignoring the black markers highlighted in cyan; these represent the five
additional NR simulations that were necessary to improve the model.
}
\label{fig:parameter_distribution}
\end{figure}

With this insight, we finally performed five new SpEC NR simulations at these
points and added the hybrid waveforms based on these to our training set for
the final model, which now includes 20 NR-EOB hybrid waveforms and 31 \NRSurOld
waveforms, for a total of 51 waveforms. Figure~\ref{fig:parameter_distribution}
shows the distribution of these parameters, including the failed simulation and
the new $q=14$ simulations.

The new NR simulations are performed using SpEC~\cite{SpECwebsite,
Boyle:2019kee}; they have been assigned identifiers SXS:BBH:2463-SXS:BBH:2482,
and made publicly available through the SXS catalog~\cite{SXSCatalog}.  The
constraint equations are solved employing the extended conformal thin sandwich
formalism~\cite{York:1998hy, Pfeiffer:2002iy} with superposed harmonic Kerr
free data~\cite{Varma:2018sqd}. The evolution equations are solved employing
the generalized harmonic formulation~\cite{Lindblom:2005qh,Rinne:2008vn}.  The
start time of these simulations is approximately $5000M$ before the peak of the
waveform amplitude (defined in Eq.~(\ref{eq:amp_tot})), where $M = m_1 + m_2$
is the total Christodoulou mass measured after the initial burst of junk
radiation~\cite{Boyle:2019kee}.  The initial orbital parameters  are chosen
through an iterative procedure~\cite{Buonanno:2010yk} such that the orbits are
quasicircular; the largest eccentricity for these simulations is
$6.4\times10^{-4}$, while the median value is $2.9\times10^{-4}$. The waveforms
are extracted at several extraction surfaces at varying finite radii form the
origin and then extrapolated to future null infinity~\cite{Boyle:2009vi}.
Finally, the extrapolated waveforms are corrected to account for the initial
drift of the center of mass~\cite{Woodford:2019tlo}.

\subsubsection{Hybridization}
\label{sec:hybridization}

\begin{figure}[thb]
\centering
\includegraphics[width=8.4cm]{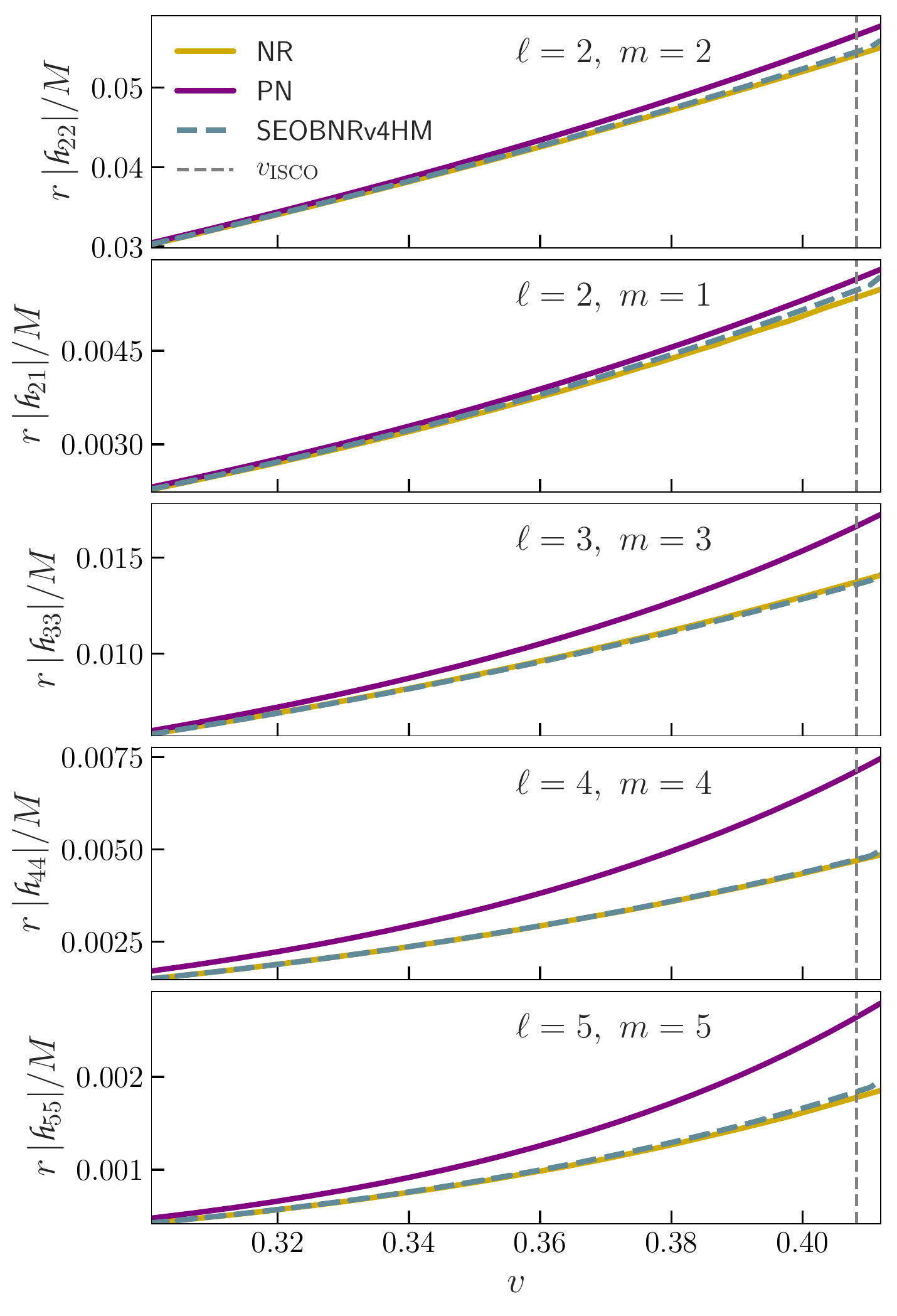}
\caption{
Mode amplitudes for NR, PN, and \EOBHM as a function of the characteristic
speed $v=\omega_{\mathrm{orb}}^{1/3}$, for binary parameters $(q, \chi_{1z},
\chi_{2z}) = (15, 0.5, 0.0)$. The vertical dashed lines represent the
Schwarzschild ISCO point $v=1/\sqrt{6}$. While PN deviates significantly from
NR, \EOBHM shows excellent agreement. We show all available modes of \EOBHM.
}
\label{fig:pn_eob_amplitudes}
\end{figure}

Given the new NR waveforms, we now hybridize them by smoothly attaching an
EOB waveform for the early inspiral. For the previous NR hybrid surrogate model
\NRSurOld~\cite{Varma:2018mmi}, a combination of PN and EOB was used for the
early inspiral: the amplitudes for all modes were obtained from PN, while the
phase evolution for all modes was derived from the (2,2) mode of the \EOBnoHM
EOB model~\cite{Bohe:2016gbl} (see Sec.~IV.B of Ref.~\cite{Varma:2018mmi}).
This was motivated by the fact that the PN mode amplitudes were found to be
accurate enough for hybridizing $q\leq8$ NR simulations, while the PN mode
phases were not (see Fig.~\ref{fig:pn_eob_amplitudes} of
Ref.~\cite{Varma:2018mmi}).

We find that the same strategy does not work for the large $q$ cases
considered in this work. Figure~\ref{fig:pn_eob_amplitudes} shows a comparison
between the mode amplitudes of NR, PN and the \EOBHM EOB
model~\cite{Cotesta:2018fcv}, for a $q=15$ system. We show all modes [(2,2),
(2,1), (3,3), (4,4), and (5,5)] included by \EOBHM, which is an extension of
the \EOBnoHM model. The PN waveforms are described in Sec.~\ref{sec:pn_sur}; we
include amplitudes terms up to 3.5 PN order~\cite{Blanchet:2008je, Faye:2012we,
Faye:2014fra}. In Fig.~\ref{fig:pn_eob_amplitudes}, the PN amplitudes
(especially for the subdominant modes) deviate significantly from NR , while
\EOBHM shows excellent agreement. This is not surprising, as \EOBHM is
calibrated to NR waveforms, as well as some BH perturbation theory waveforms at
extreme mass ratios~\cite{Cotesta:2018fcv}. We conclude that current PN
waveforms are not suitable for hybridizing NR waveforms at large mass ratios
like $q\sim 15$.  Therefore, in this work, we only use \EOBHM for hybridizing
NR waveforms. Unfortunately, this means that our new model \NRSurNew is
restricted to the same set of modes as \EOBHM.

We follow the same hybridization procedure as Sec.~V of
Ref.~\cite{Varma:2018mmi} to smoothly attach \EOBHM inspirals to the 20 new
$q>8$ NR simulations obtained in Sec.~\ref{sec:pn_sur}. For the remaining 31
training cases with $q\leq8$, we generate waveforms using the \NRSurOld model,
as it is already hybridized. This completes the construction of our training
set waveforms.

\subsection{Frame alignment}

We follow Ref.~\cite{Varma:2018mmi} and apply the following post processing to
the training set waveforms. This ensures that all waveforms are in the same
frame, and therefore that the data used in the surrogate fits (see
Sec.~\ref{sec:surrogate_building}) vary smoothly across parameter space.

\subsubsection{Time alignment}
\label{sec:time_shifts}
We apply a time shift to each training waveform such that peak of the total
amplitude
\begin{equation}
    A_{\mathrm{tot}} = \sqrt{\sum_{l,m} \lvert\hlm\rvert^2},
\label{eq:amp_tot}
\end{equation}
occurs at $t=0$. The original peak time is determined by a quadratic fit using
5 time samples adjacent to the discrete maximum of
$A_{\mathrm{tot}}$~\cite{Blackman:2017pcm}.

\subsubsection{Down-sampling and common time array}
\label{sec:common_time_array}

The length of each hybrid waveform is set by choosing a starting orbital
frequency $\omega_{\mathrm{orb}}$ for the \EOBHM inspiral; we use
$\omega_{\mathrm{orb}} = 1 \times 10^{-3} \, \mathrm{rad}/M$ for all waveforms.
However, for the same starting frequency, the waveform length in time is
different for different mass ratios and spins. On the other hand, the
surrogate modeling procedure requires that all training waveforms have a common
time array~\cite{Field:2013cfa}. Therefore, we truncate all waveforms such that
they start at the same initial time ($\sim 2.4 \times 10^7 \,M$ before the
peak), which is determined by the shortest hybrid waveform in the training set.
Post truncation, the largest starting orbital frequency is
$\omega_{\mathrm{orb}} = 1.1 \times 10^{-3} \, \mathrm{rad}/M$, which sets the
low-frequency limit of validity of the surrogate. For LIGO and Virgo, assuming a starting
GW frequency of $20 \mathrm{Hz}$, the $(2,2)$ mode of the surrogate model is
valid for total masses $M \geq 3.7 \, M_{\odot}$. The highest spin-weighted
spherical harmonic mode included in the model is $(5,5)$, for which the
corresponding frequency is 5/2 times that of the $(2,2)$ mode. Therefore, all
modes of the surrogate are valid for $M\gtrsim 9.5 \, M_{\odot}$.

Because the hybrid waveforms are very long, it is not practical to sample the
entire waveform with a small uniform time step like $0.1\,M$, as is typically
done for NR-only surrogates~\cite{Varma:2019csw}. Fortunately, the early
low-frequency portion of the waveform does not require as dense a time sampling
as the later high-frequency portion. We therefore down-sample the time arrays
of the truncated hybrid waveforms to a common set of time samples. We choose
the time samples such that there are $5$ points per orbit for the
above-mentioned shortest hybrid waveform in the training set. However, for $t
\geq -1000\, M_{\odot}$ we switch to uniformly spaced time samples with a time
step of $0.1\,M$. This ensures that we have a sufficiently dense sampling rate
for the late inspiral and the merger-ringdown where the frequency reaches its
peak. We retain times up to $120 \, M$ after the peak, which is sufficient to
capture the entire ringdown.

Given the common down-sampled time array, we use cubic splines to interpolate
all waveforms in the training set to these times. However, we first transform
the waveforms into the co-orbital frame, defined as:
\begin{gather}
    \hlm^C = \hlm \, e^{im\phi_{\mathrm{orb}}} \label{eq:coorbital}, \\
    \h_{22} = A_{22} \, e^{-i\phi_{22}},
    \label{eq:h22}\\
    \phi_{\mathrm{orb}} = \frac{\phi_{22}}{2},
    \label{eq:orb_phase}
\end{gather}
where $\hlm$ is the inertial frame waveform, $\phi_{\mathrm{orb}}$ is the
orbital phase, and $A_{22}$ and $\phi_{22}$ are the amplitude and phase of the
$(2,2)$ mode. The co-orbital frame can be seen as roughly co-rotating with the
binary, obtained by applying a time-dependent rotation about the $z-$axis, by
an amount given by the instantaneous orbital phase. Therefore, the waveform is
a slowly varying function of time in this frame, which increases the
interpolation accuracy. For the (2, 2) mode we save the downsampled amplitude
$A_{22}$ and phase $\phi_{22}$, while for all other modes we save the real and
imaginary parts of $\hlm^C$.

\subsubsection{Phase alignment}
\label{sec:phase_alignment}
Finally, we rotate the waveforms about the $z$-axis such that the orbital phase
$\phi_{\mathrm{orb}}$ is zero at $t=-1000\,M$. Note that this by itself would
fix the physical rotation up to a shift of $\pi$. When generating the EOB
inspiral waveform for hybridization, the frame is aligned such that heavier BH
is on the positive $x$-axis at the initial time, which fixes the $\pi$
ambiguity~\cite{Varma:2018mmi}. After the phase alignment, the heavier BH is
on the positive $x$-axis at $t=-1000\,M$ for all waveforms. However, keep in
mind that this frame is defined using the waveform at future null infinity, and
these BH positions do not necessarily correspond to the (gauge-dependent)
coordinate BH positions in the NR simulations.

\subsection{Data decomposition}
\label{sec:decomposing_the_data}

It is much easier to build a model for slowly varying functions of time.
Therefore, we decompose the inertial frame strain $\hlm$, which is oscillatory,
into simpler ``waveform data pieces'' and build a separate surrogate for each
data piece. When evaluating the full surrogate model, we first evaluate the
surrogate for each data piece and then combine the data pieces to get the
inertial frame strain. The $(2,2)$ mode is decomposed into its amplitude
$A_{22}$ and phase $\phi_{22}$ (which is further decomposed below). For the
other modes, we model the real and imaginary parts of the co-orbital frame
strain $\hlm^C$ (see Eq.~(\ref{eq:coorbital})).

Following Ref.~\cite{Varma:2019csw}, we further decompose $\phi_{22}$ by
subtracting the leading-order prediction from the TaylorT3 PN
approximant~\cite{Damour:2000zb}, given by:
\begin{equation}\label{eq:T3}
    \phi_{22}^{\mathrm{T3}} = \phi_{\mathrm{ref}}^{\mathrm{T3}} - \frac{2}{\eta \theta^5},
\end{equation}
where $\phi_{\mathrm{ref}}^{\mathrm{T3}}$ is an arbitrary integration constant,
$\theta = [ \eta (t_{\mathrm{ref}}-t)/(5M)]^{-1/8} $, $t_\mathrm{ref}$ is an
arbitrary time offset, and $\eta=q/(1+q)^2$ is the symmetric mass ratio.
Because $\phi_{\mathrm{ref}}^{\mathrm{T3}}$ diverges at $t_\mathrm{ref}$, we
choose $t_\mathrm{ref}=1000\,M$, long after the peak ($t=0$) of the waveform,
ensuring that we are always far away from this divergence. We choose
$\phi_{\mathrm{ref}}^{\mathrm{T3}}$ such that $\phi_{22}^{\mathrm{T3}} = 0$ at
$t=-1000\,M$, which is the same time at which we align the hybrid phase in
Sec.~\ref{sec:phase_alignment}.

By modeling the difference $\phi_{22}^\mathrm{res} = \phi_{22} -
\phi_{22}^{\mathrm{T3}}$ instead of $\phi_{22}$, we automatically capture
almost all of the phase evolution in the early inspiral of the long hybrid
waveforms.  Therefore, we simplify the problem of modeling the phase to the
same as modeling the phase of NR-only waveforms. This improves the overall
accuracy of the surrogate model for low masses, for which the inspiral
dominates. We stress that the exact form of $\phi_{22}^{\mathrm{T3}}$ (or its
physical meaning) is not important because we add the exact same
$\phi_{22}^{\mathrm{T3}}$ to our model of $\phi_{22}^\mathrm{res}$ when
evaluating the surrogate. In fact, even though TaylorT3 is known to be less
accurate than other approximants~\cite{Buonanno:2009zt, Varma:2013kna}, its
speed (being a simple, analytic, closed-form, function of time) makes it ideal
for our purpose.

To summarize, we decompose the hybrid waveforms into the following waveform
data pieces, each of which is a smooth, slowly varying function of time:
$(A_{22}, \phi_{22}^\mathrm{res})$ for the $(2,2)$ mode, and the real and
imaginary parts of $\hlm^C$ for the (2,1), (3,3), (4,4) and (5,5) modes.

\begin{figure*}[tbh]
\centering
\includegraphics[width=0.49\textwidth]{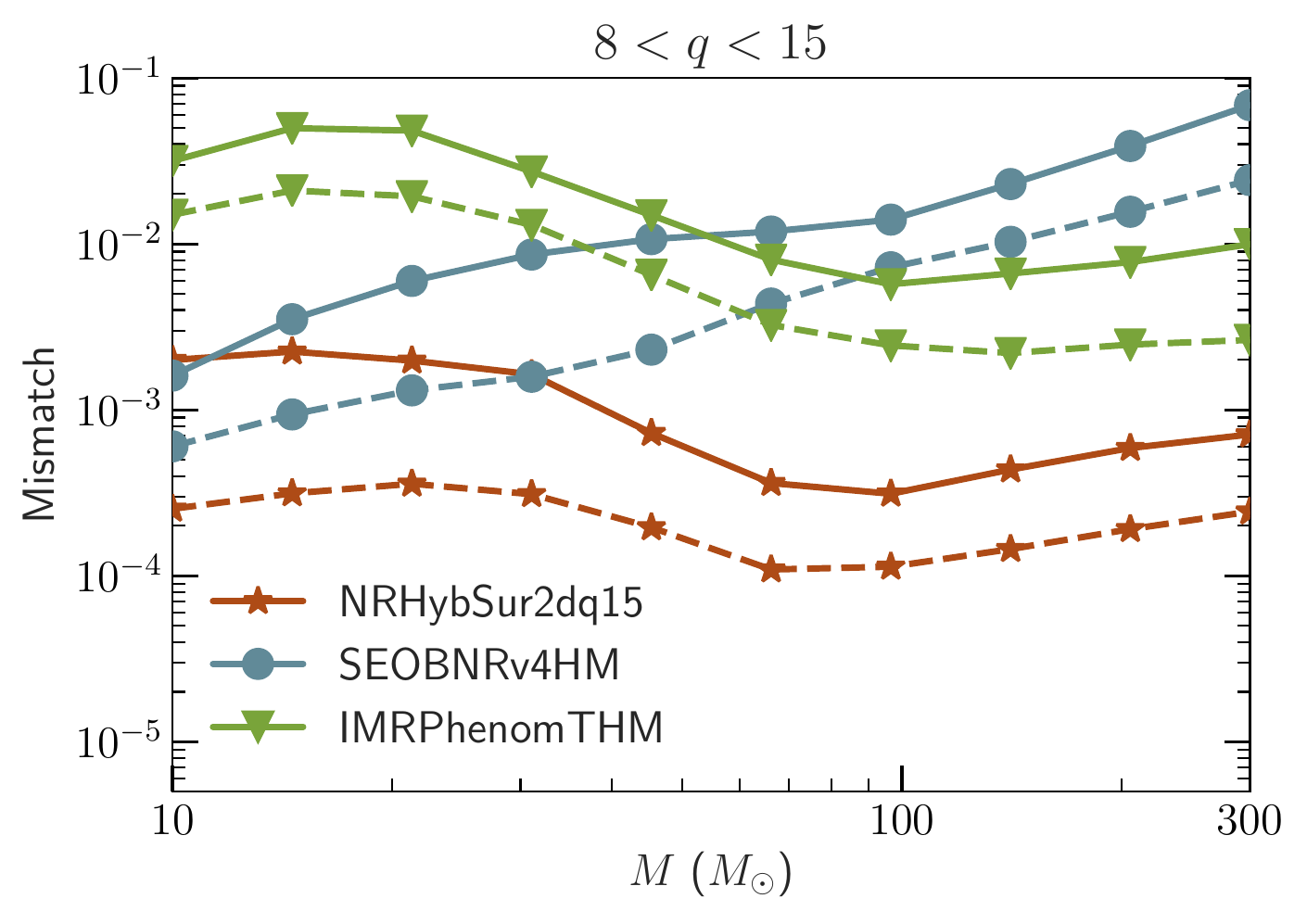}
\includegraphics[width=0.49\textwidth]{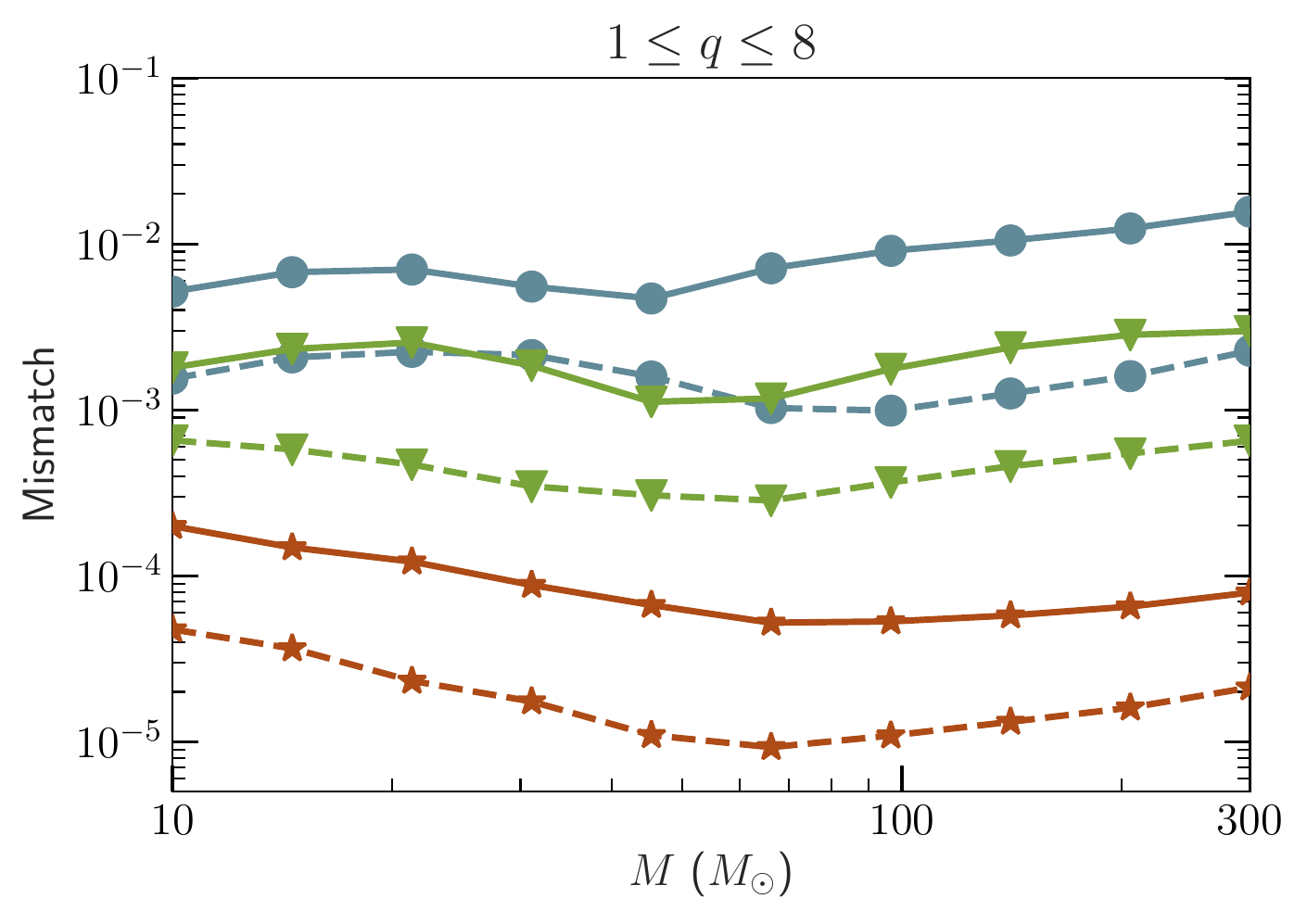}
\caption{
\emph{Left:} Mismatches as a function of the total mass M for \NRSurNew,
\EOBHM and \IMRTHM against NR-EOB hybrid waveforms with $q>8$. For \NRSurNew,
we show leave-one-out errors. Mismatches are computed using the Advanced-LIGO
noise curve, at several points in the sky of the source frame using all
available modes: (2,2), (2,1), (3,3), (4,4), and (5,5).  The solid (dashed)
lines show the 95th percentile (median) mismatch values over points on the sky
as well as different hybrid waveforms.  \emph{Right:} Same, but now the
mismatches are computed against the \NRSurOld model in the $q\leq8$ region.
}
\label{fig:ligo_mis}
\end{figure*}

\subsection{Surrogate construction and evaluation}
\label{sec:surrogate_building}
Given the waveform data pieces, we build a surrogate model for each data piece
using the same procedure as Sec.V.C of Ref.~\cite{Varma:2018mmi}, which we
summarize below.

For each waveform data piece, we first construct a linear basis using the
greedy basis method~\cite{Field:2011mf}, with tolerances of $10^{-2}$ radians
for the $\phi_{22}^{\mathrm{res}}$ data piece and $5 \times 10^{-5}$ for all
other data pieces. Next, we construct an empirical time
interpolant~\cite{Barrault:2004, Maday:2009, Hesthaven:2014} with the same
number of empirical time nodes as basis functions for that data piece.
Finally, for each empirical time node, we construct a parametric fit for the
waveform data piece, following the Gaussian process regression (GPR) fitting
method, as described in Refs.~\cite{Varma:2018aht, Taylor:2020bmj}. The fits
are parameterized by $(\log(q), \hat{\chi})$, where
\begin{gather}
\hat{\chi}  = \frac{\chieff-38\eta(\chi_{1z}+\chi_{2z})/113}{1-76\eta/113},
\end{gather}
is the spin parameter entering the GW phase at leading
order~\cite{Ajith:2011ec}, and $\chieff = \frac{q \chi_{1z} + \chi_{2z}}{1+q}$
is the effective spin. Note that in the above expressions $\chi_{2z}=0$ for the
current surrogate, but we adopt this parameterization to be consistent with
Ref.~\cite{Varma:2018mmi}. In practice, parameterizing the fits by $(\log(q),
\chi_{1z})$ also leads to a surrogate of similar accuracy.  On the other hand,
the $\log(q)$ parameterization leads to a significant improvement in model
accuracy, in agreement with Refs.~\cite{Varma:2018mmi, Rifat:2019ltp}.

When evaluating the surrogate waveform, we first evaluate each surrogate
waveform data piece. Next, we compute the $(2,2)$ mode phase:
\begin{equation}
    \phi_{22}^{\mathrm{S}} \equiv \phi_{22}^{\mathrm{res,S}} +
    \phi_{22}^{\mathrm{T3}},
\end{equation}
where $\phi_{22}^{\mathrm{res,S}} \approx \phi_{22}^{\mathrm{res}}$ is the
surrogate model for $\phi_{22}^{\mathrm{res}}$, and $ \phi_{22}^{\mathrm{T3}}$
is given by Eq.~(\ref{eq:T3}). If the waveform is required at a uniform
sampling rate, we interpolate each waveform data piece from the sparse time
samples to the required time samples using a cubic-spline interpolation scheme.
Finally, we use Eqs.~(\ref{eq:coorbital}),~(\ref{eq:h22}),
and~(\ref{eq:orb_phase}) to reconstruct the inertial frame strain.

\section{Surrogate errors}
\label{sec:surrogate_errors}

In this section, we evaluate the accuracy of \NRSurNew by comparing against
NR-EOB hybrid waveforms. Similarly, we compute errors for two semi-analytic
waveform models, the phenomenological model \IMRTHM~\cite{Estelles:2020twz} and
the EOB model \EOBHM~\cite{Cotesta:2018fcv}. Both of these models are
calibrated against nonprecessing NR simulations and include the same set of
modes as \NRSurNew and the hybrid waveforms: (2,2), (2,1), (3,3), (4,4) and
(5,5).  Other semi-analytic nonprecessing models that include subdominant modes
exist in literature, including Refs~\cite{Nagar:2020pcj,
Garcia-Quiros:2020qpx}, but we do not consider these models for simplicity (as
they have accuracies comparable~\cite{Nagar:2020pcj, Garcia-Quiros:2020qpx,
Nagar:2022icd} to \IMRTHM and \EOBHM).

In order to estimate the difference between two waveforms, $\h_1$ and $\h_2$,
we compute the mismatch (Eq.~\ref{eq:mismatch_td}) using the
noise-weighted inner product in frequency-domain, defined as
\begin{equation}
\braket{\h_1,\h_2} = 4 \Re \left[ \int_{f_{\mathrm{min}}}^{f_{\mathrm{max}}} \frac{\tilde{\h}_1(f)\tilde{\h}^*_2(f)}{S_n(f)} df \right],
\label{eq:mismatch_fd}
\end{equation}
where $\tilde{\h}(f)$ indicates the Fourier transform of the complex strain
$\h(t)$, $^*$ indicates a complex conjugation, $\Re$ indicates the real part,
and $S_n(f)$ is the one-sided power spectral density of a GW detector. We use
the Advanced-LIGO design sensitivity Zero-Detuned-HighP noise
curve~\cite{aLIGODesignNoiseCurve}, with $f_{\mathrm{min}} = 20$ Hz and
$f_{\mathrm{max}} = 2000$ Hz. We compute the mismatches following the
procedure described in Sec.VII of Ref.~\cite{Varma:2018mmi}: the mismatches are
optimized over shifts in time, polarization angle, and initial orbital phase.
Both plus and cross polarizations are treated on an equal footing by using a
two-detector setup where one detector sees only the plus and the other only the
cross polarization.  We use all the available modes of a given waveform model,
and compute the mismatches at 37 points uniformly distributed on the sky in the
source frame.

Figure~\ref{fig:ligo_mis} shows mismatches computed using the Advanced-LIGO
noise curve for \NRSurNew, \EOBHM and \IMRTHM against hybrid waveforms. As
these depend on the total mass, we show mismatches for various masses, starting
near the lower limit of the range of validity of the surrogate $M\gtrsim 9.5 \,
M_{\odot}$. At each mass, we show the median and 95th percentile mismatches,
over many hybrid waveforms and points in the source frame sky.

The left panel of Fig.~\ref{fig:ligo_mis} shows mismatches against the 20 $q>8$
NR-EOB hybrid waveforms in Fig.~\ref{fig:parameter_distribution}. As these
hybrid waveforms were also used in the training of \NRSurNew, we conduct a
\emph{leave-one-out} analysis: we generate 20 trial surrogates, leaving out one
of the $q>8$ hybrid waveforms from the training set in each trial, but
including the rest of the training cases (both $q>8$ and $q\leq8$) in
Fig.~\ref{fig:parameter_distribution}. For each trial surrogate, we compute
errors against the $q>8$ hybrid waveform that was left out. In this manner, we
only compare \NRSurNew against waveforms not used in the model training.
Therefore, these errors are indicative of the true modeling error.

For the $q>8$ region, 95th percentile mismatches for \NRSurNew fall below $\sim
2\times10^{-3}$ over the entire mass range in Fig.~\ref{fig:ligo_mis}. The
errors for \IMRTHM and \EOBHM are generally larger by at least an order of
magnitude.  However, for \EOBHM, the errors at low masses overlap with the
surrogate errors. This is most likely because \EOBHM was used to generate the
early inspiral waveform for the NR-EOB hybrid waveforms. At low masses, where
the early inspiral dominates the overall error budget, these errors are
therefore not representative of the true error in \EOBHM.

The right panel of Fig.~\ref{fig:ligo_mis} shows mismatches in the $q<8$
region. In this region, rather than conduct leave-one-out tests, we simply
generate 100 new hybrid waveforms using the \NRSurOld model for testing.  These
test cases are uniformly distributed in the region $q\in[1,8]$ and
$\chi_{1z}\in[-0.5, 0.5]$, with $\chi_{2z}=0$. Once again \NRSurNew has
mismatches that are at least an order of magnitude smaller than that of \EOBHM
and \IMRTHM. In this case, \EOBHM errors are broadly uniform across all masses.
This is most likely explained by the fact that the early inspiral of \NRSurOld
was based on PN as well as EOB waveforms; more precisely, PN was directly used
to generate the mode amplitudes while the (2,2) mode of \EOBHM (the
\EOBnoHM~\cite{Bohe:2016gbl} model) was used to correct the PN mode phases.

\begin{figure*}[h]
\centering
\includegraphics[width=0.45\textwidth]{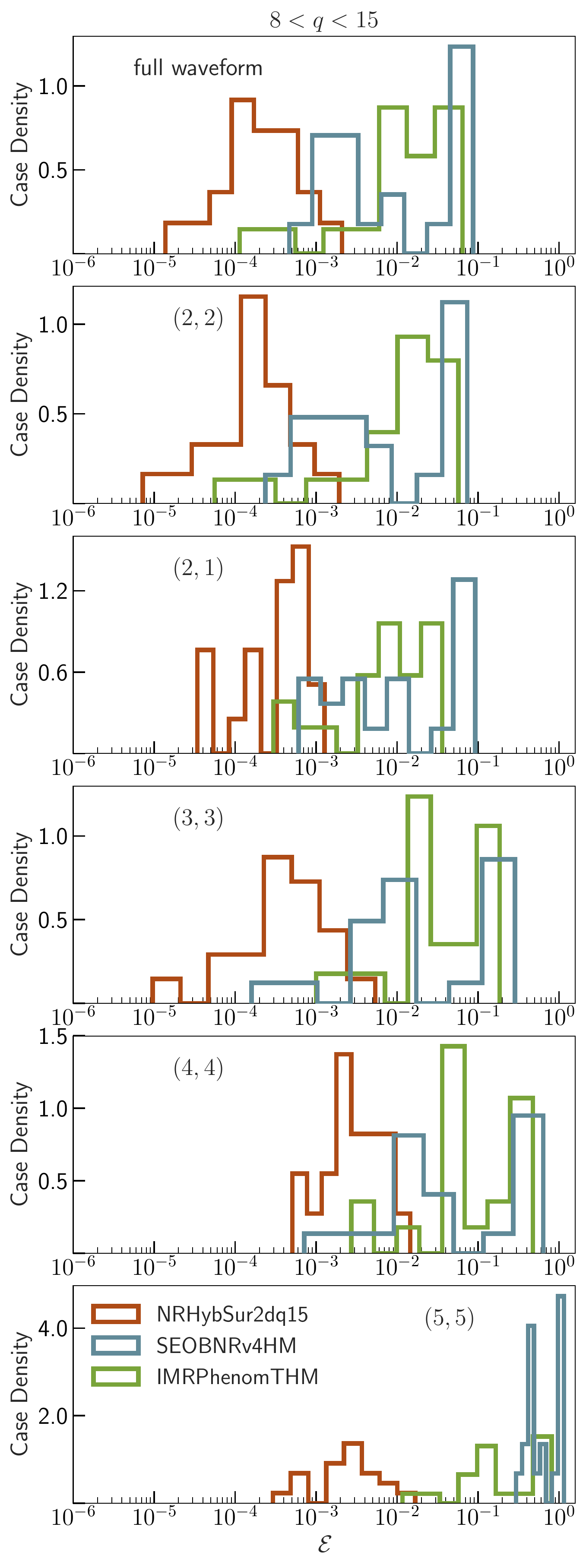}
\includegraphics[width=0.45\textwidth]{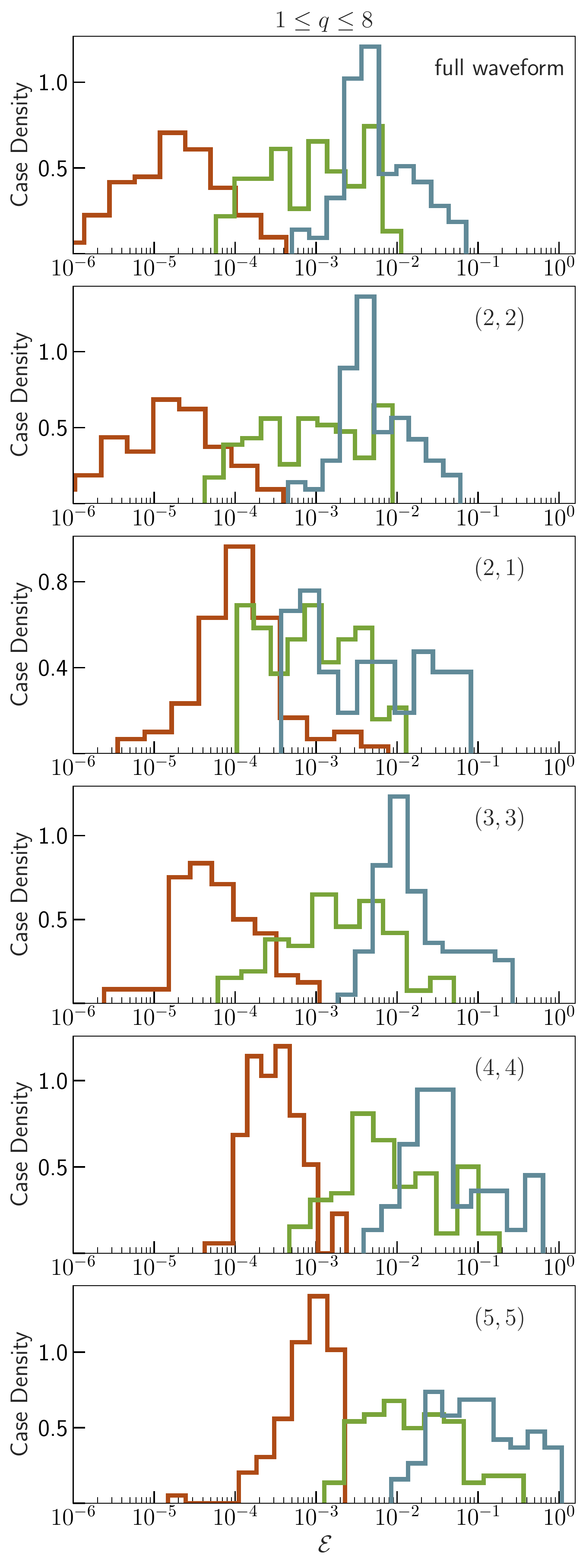}
\caption{
\emph{Left:} Normalized error, $\mathcal{E}$ (Eq.~(\ref{eq:L2_error})),
computed for \NRSurNew, \EOBHM, and \IMRTHM against NR-EOB hybrid waveforms
with q>8, but restricting the start time of the waveforms to $-4500M$ before
the peak amplitude. In the first row, $\mathcal{E}$ is computed using all
available modes, and in the subsequent rows, single-mode errors are computed by
restricting Eq.~(\ref{eq:L2_error}) to individual modes. \emph{Right:} Same,
but now the error is computed against the \NRSurOld model in the $q\leq8$
region.
}
\label{fig:error_histgoram}
\end{figure*}

\begin{figure*}[tbh]
\centering
\includegraphics[width=\textwidth]{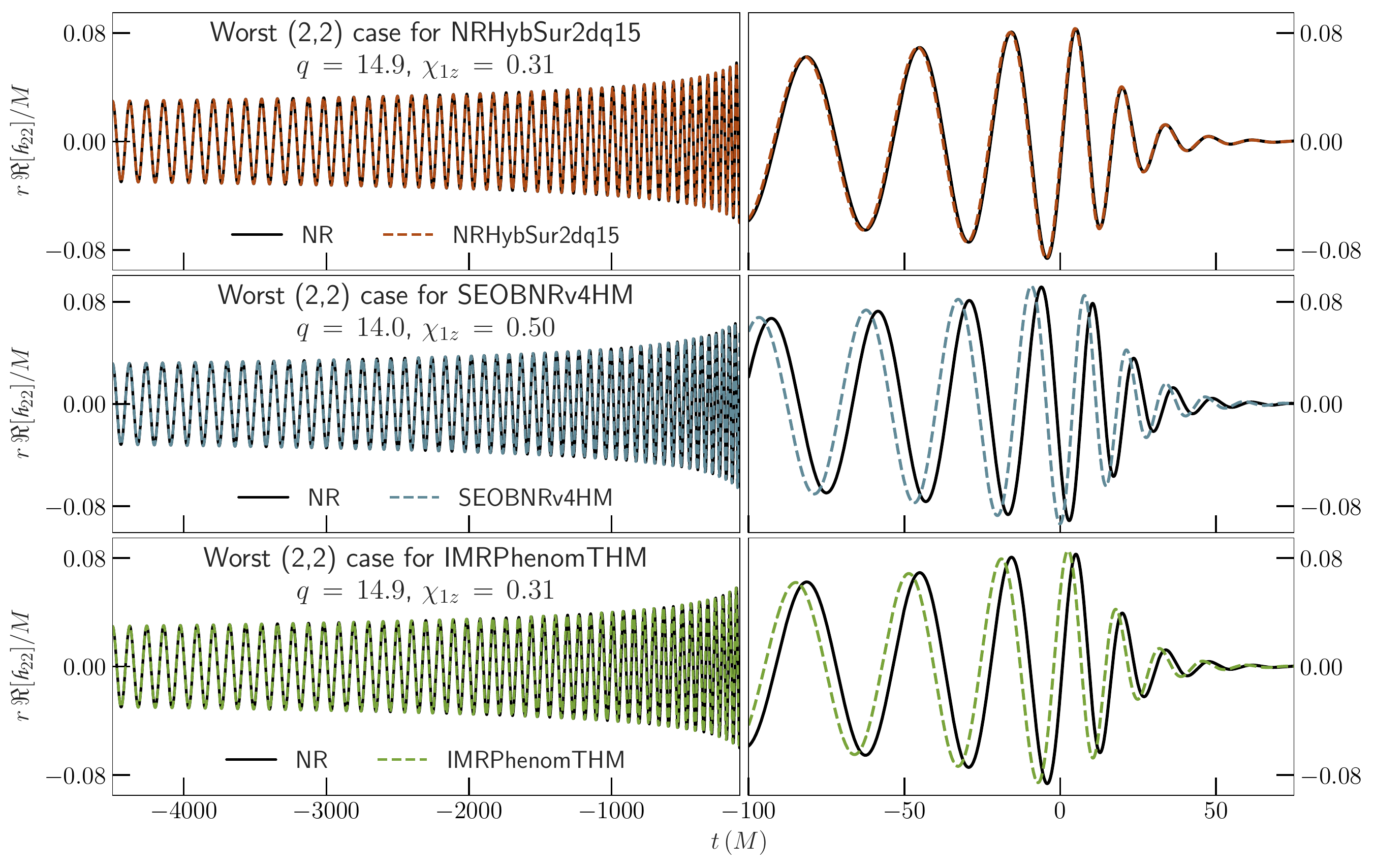}
\caption{
The (2,2) modes of the three waveform models compared against NR, for the cases
that lead to the largest (2,2) mode error in the left panel of
Fig.~\ref{fig:error_histgoram}. The top (middle) [bottom] panel shows the case
for which \NRSurNew (\EOBHM) [\IMRTHM] has the largest (2,2) mode error.
}
\label{fig:WorstCasePlot_22}
\end{figure*}

While Fig.~\ref{fig:ligo_mis} shows model errors when including all available
modes, it can be useful to also understand the errors in the individual modes.
We quantify this using the normalized $L_2-$norm between two waveforms $\h$ and
$\h'$:
\begin{align}
    \mathcal{E}(\h, \h') = \frac{1}{2}\frac{\sum_{l,m}\int^{t_2}_{t_1}\vert \hlm(t) - \hlm^{'}(t) \vert^2 dt}{\sum_{l,m}\int^{t_2}_{t_1}\vert \hlm(t)\vert^2 dt}.
\label{eq:L2_error}
\end{align}
This error measure was introduced in Ref.~\cite{Blackman:2017dfb} and is
related to weighted average of the mismatch over the sky in the source frame.
When computing $\mathcal{E}$, we only consider the late inspiral and
merger-ringdown region by choosing $t_1=-4500M$ and $t_2=115M$. As the NR
waveforms used in generating the hybrid waveforms had typical start times $\sim
-5000M$ (see Sec.~\ref{sec:training_set}), this ensures that $\mathcal{E}$ is
independent of which model was used in the hybridization procedure.
Furthermore, rather than optimizing over time or phase shifts, we simply align
the frames of the two waveforms such that the peak amplitude
(Eq.~(\ref{eq:amp_tot})) occurs at $t=0$, and the orbital phase
(Eq.~(\ref{eq:orb_phase})) is zero at $t = -4500M$.  This makes $\mathcal{E}$
much cheaper to evaluate than the mismatches in Eq.~(\ref{eq:mismatch_fd}).  In
addition to computing normalized errors using all available modes, we also
consider single-mode errors by restricting the sums in Eq.~(\ref{eq:L2_error})
to individual modes.

Figure~\ref{fig:error_histgoram} shows normalized errors for \NRSurNew, \EOBHM
and \IMRTHM against hybrid waveforms. The left panel of
Fig.~\ref{fig:error_histgoram} follows the left panel of
Fig.~\ref{fig:ligo_mis}, and shows errors for the three waveform models (using
a leave-one-out analysis for \NRSurNew) against the 20 $q>8$ NR-EOB hybrid
waveforms. The right panel of Fig.~\ref{fig:error_histgoram} follows the right
panel of Fig.~\ref{fig:ligo_mis}, and shows errors against the same 100
uniformly distributed \NRSurOld waveforms in the region $q\in[1,8]$ and
$\chi_{1z}\in[-0.5, 0.5]$, with $\chi_{2z}=0$. For both $q>8$ and $q\leq8$, we
once again find that \NRSurNew is more accurate than the other models by at
least an order of magnitude, both for the full waveform and for the individual
modes.

\begin{figure*}[p]
\centering
\includegraphics[width=\textwidth]{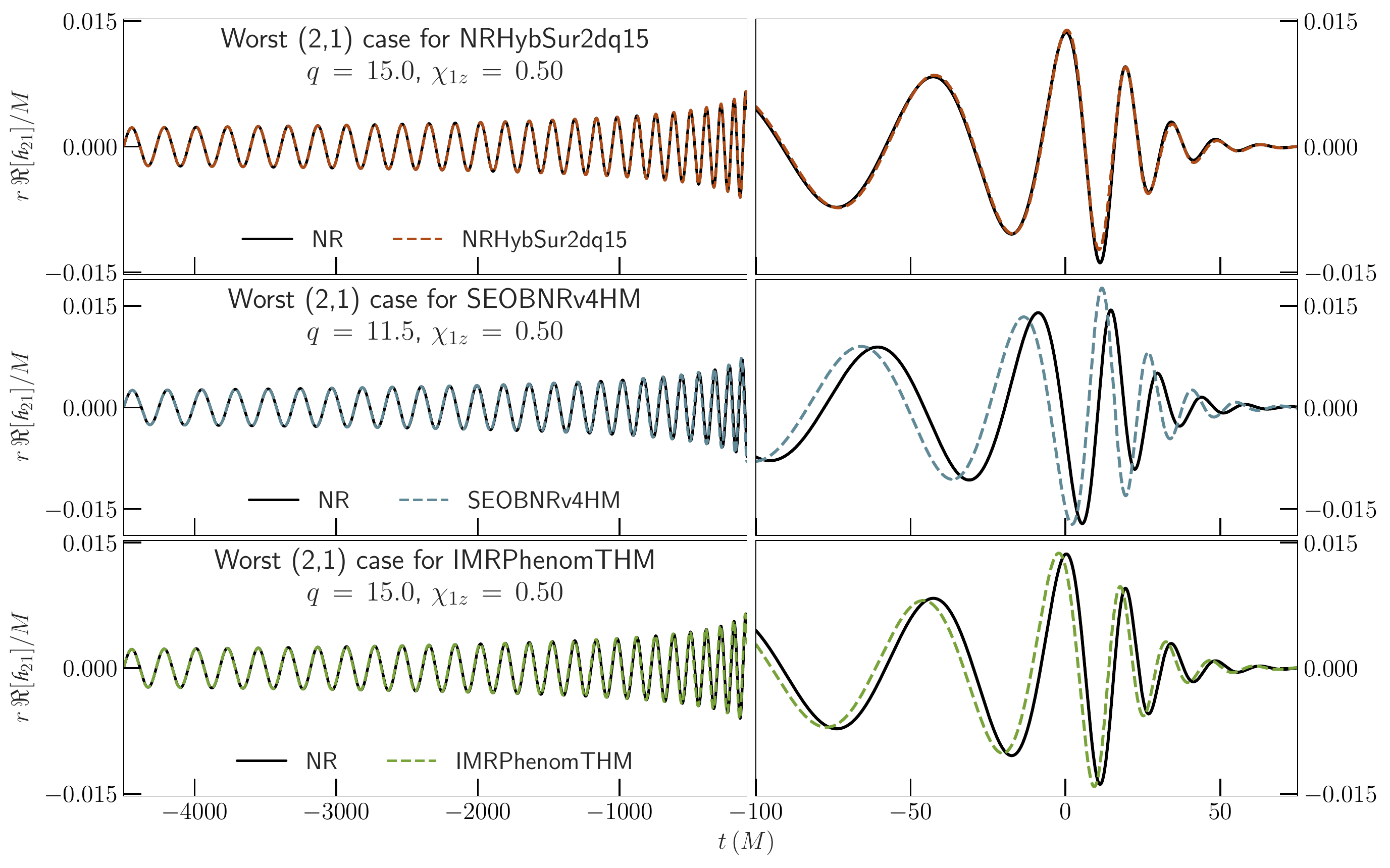}

\vspace{0.6cm}

\includegraphics[width=\textwidth]{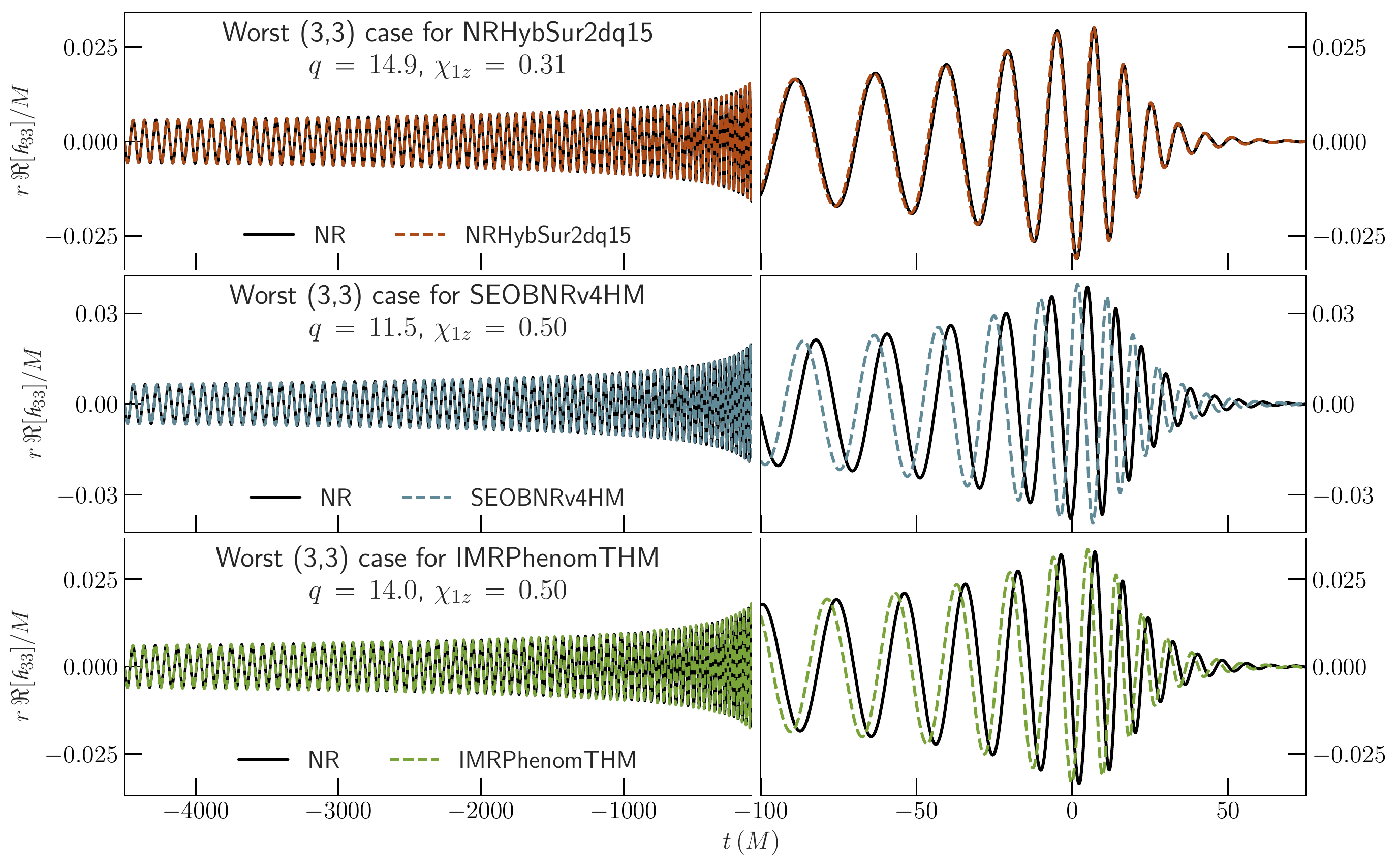}
\caption{
Same as Fig.~\ref{fig:WorstCasePlot_22}, but now showing the worst cases for
the (2,1) [top] and (3,3) [bottom] modes.
}
\label{fig:WorstCasePlot_21_33}
\end{figure*}

\begin{figure*}[p]
\centering
\includegraphics[width=\textwidth]{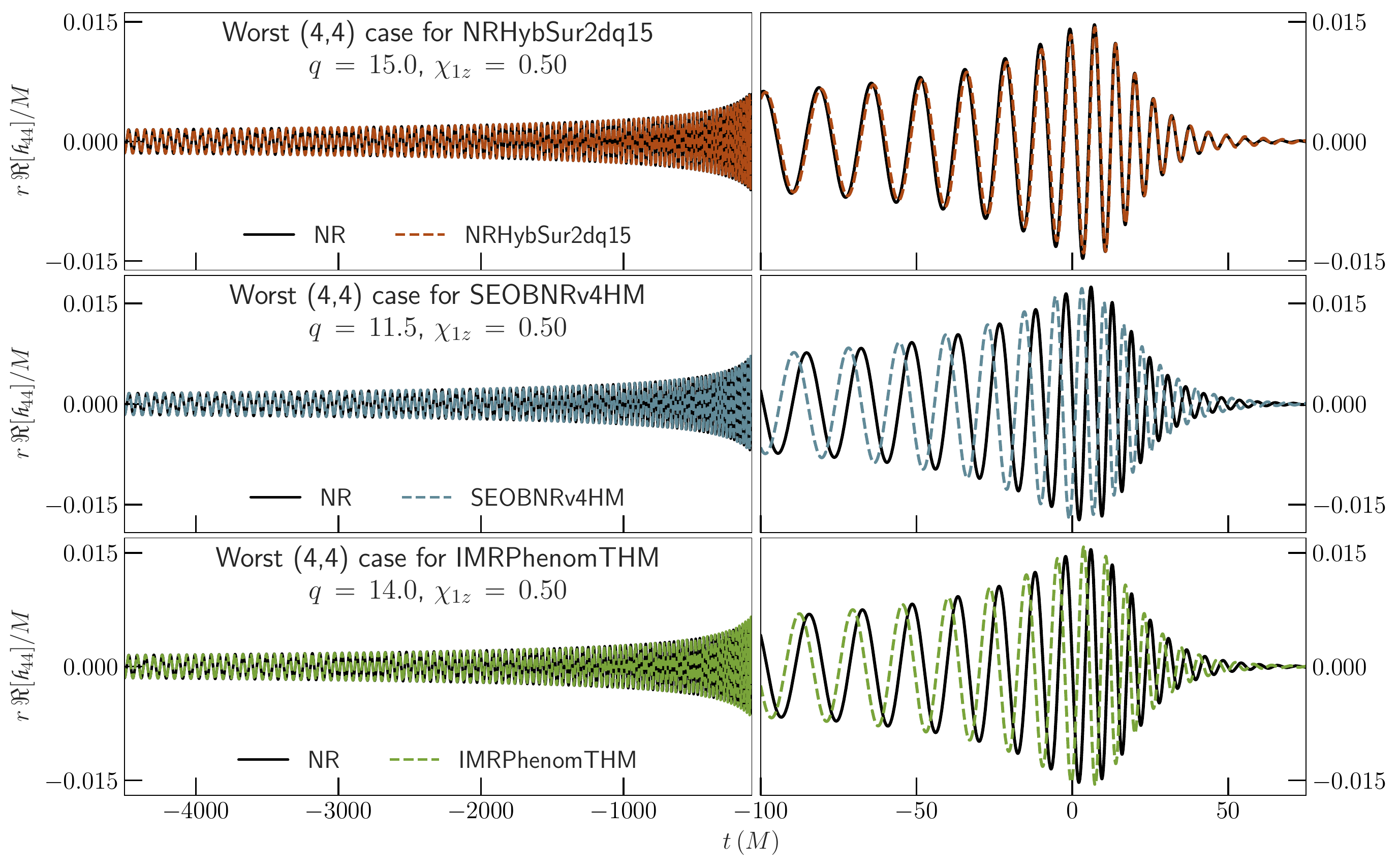}

\vspace{0.6cm}

\includegraphics[width=\textwidth]{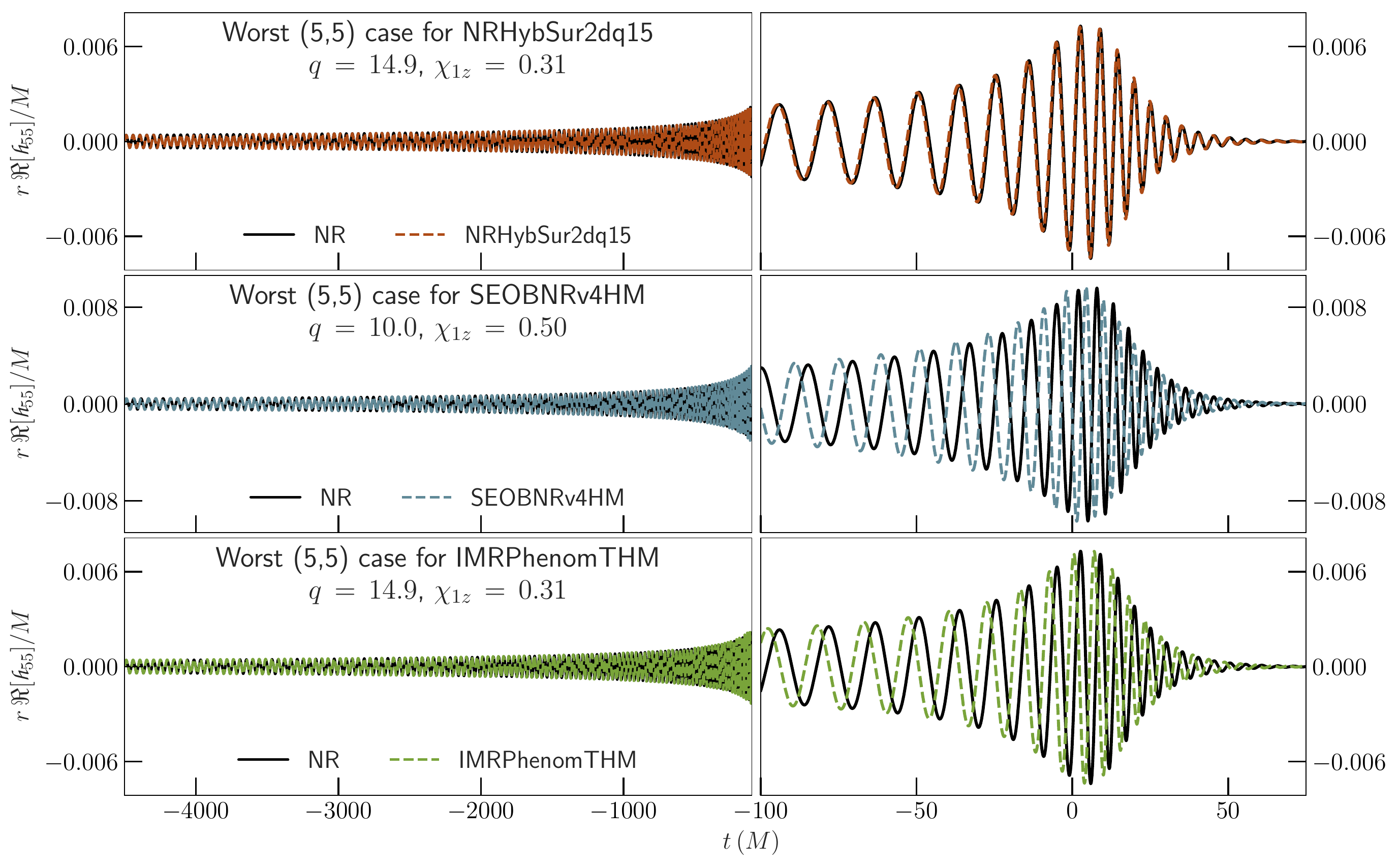}
\caption{
Same as Fig.~\ref{fig:WorstCasePlot_22}, but now showing the worst cases for
the (4,4) [top] and (5,5) [bottom] modes.
}
\label{fig:WorstCasePlot_44_55}
\end{figure*}

Considering the individual mode errors in Fig.~\ref{fig:error_histgoram}, we
note that the fractional errors in the nonquadrupole modes of \EOBHM and
\IMRTHM reach large values. In particular, the errors in the (5,5) mode for
\EOBHM for $q>8$ can reach values $\mathcal{E} \sim 1$. While the nonquadrupole
modes are still subdominant for $q\gtrsim10$ binaries like GW190814 (which is
why the full waveform errors do not reach such large values in
Fig.~\ref{fig:error_histgoram}), it may be important for models like \IMRTHM
and \EOBHM to improve accuracy in these modes for future observations. Finally,
to illustrate the (in)accuracy of the individual modes,
Figs.~\ref{fig:WorstCasePlot_22}, \ref{fig:WorstCasePlot_21_33} and
\ref{fig:WorstCasePlot_44_55} show the cases leading to the largest individual
mode errors in the left panel of Fig.~\ref{fig:error_histgoram}.

\subsection{Extrapolating outside the training region}
The errors computed so far were restricted to the training region of
\NRSurNew: $q\leq 15$, $\chi_{1z} \in [-0.5, 0.5]$, and $\chi_{2z}=0$.  It is
possible to extrapolate the model to larger $q$ and $|\chi_{1z}|$, but it is
difficult to assess the model accuracy in this region due to a lack of NR
simulations. Instead, through a visual inspection of the evaluated waveforms,
we find that extrapolating beyond $q=20$ or $|\chi_{1z}|=0.7$ leads to
unphysical ``glitches'' in the time series for the mode amplitudes and the
derivatives of the mode phases. Therefore, while we allow the model to be
evaluated in the region $q\leq20$, $\chi_{1z} \in [-0.7, 0.7]$, and
$\chi_{2z}=0$, we advise caution when extrapolating the model.

\begin{figure*}[htb]
\centering
\includegraphics[width=0.49\textwidth]{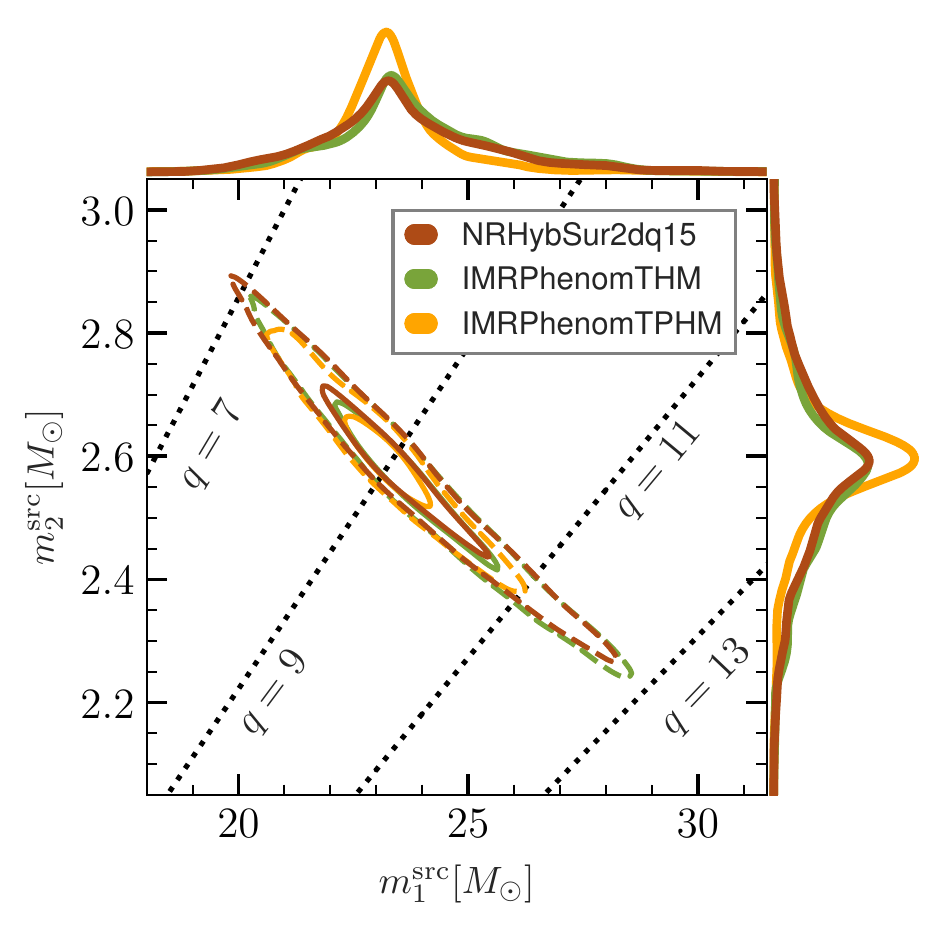}
\includegraphics[width=0.49\textwidth]{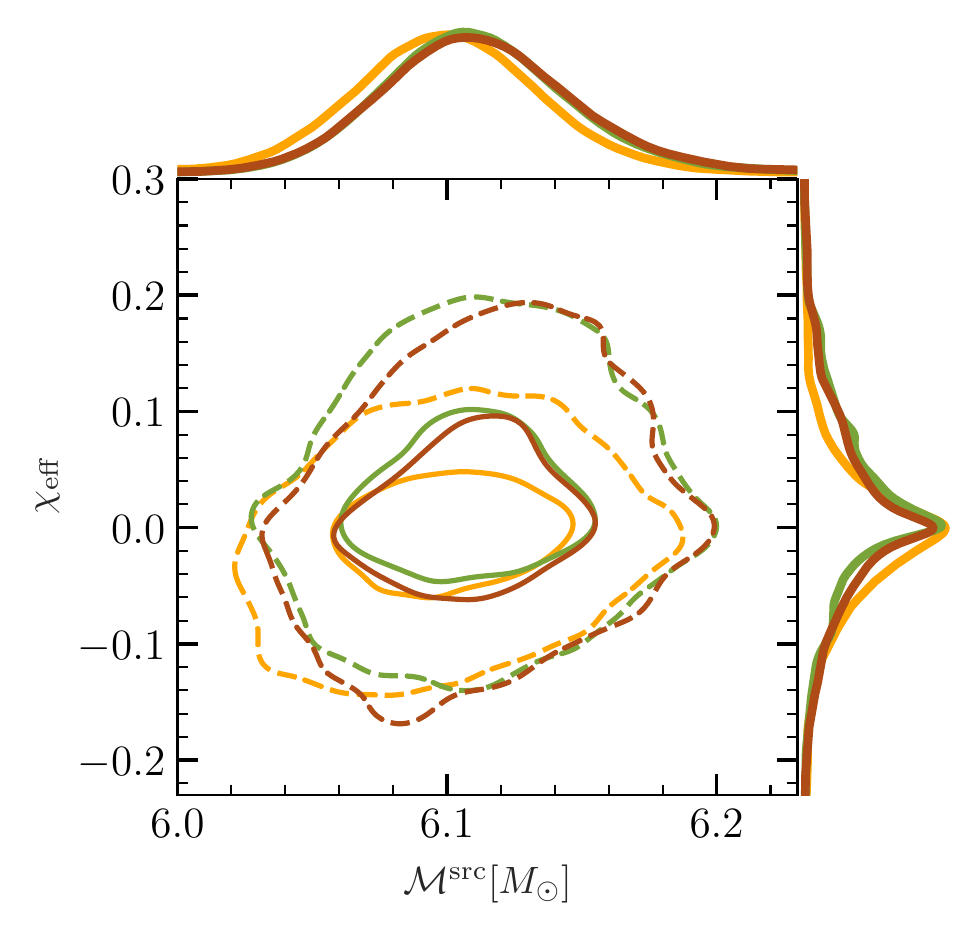}
\centering
\includegraphics[width=0.49\textwidth]{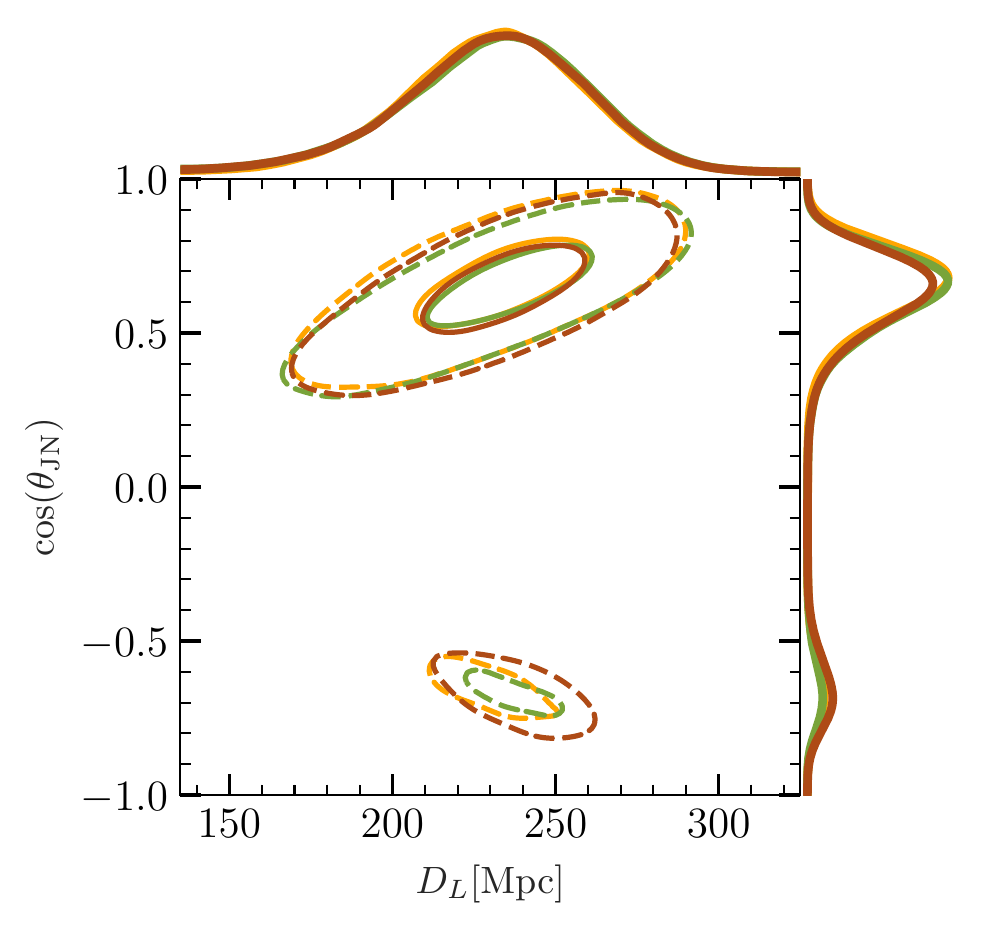}
\caption{
Constraints on GW190814 parameters obtained using the \NRSurNew, \IMRTHM and
\IMRPHMOld models. We show posterior distributions for the source-frame
component masses $\mSrc{1}$ and $\mSrc{2}$ (top-left), the effective spin
$\chieff$ and the source-frame chirp mass $\MchirpSrc$ (top-right), and the
extrinsic parameters $cos(\theta_{JN})$ and luminosity distance $D_L$ (bottom).
The solid (dashed) contours represent the central 50\% (90\%) credible regions
of the joint posteriors. Marginalized 1D posteriors are shown on the plot
edges. In the top-left panel, we include lines of constant mass ratios
($q=7,9,11,13$) for comparison. The bimodality in the bottom panel is due to a
well known degeneracy between distance and inclination~\cite{Cutler:1994ys}.
\IMRTHM and \NRSurNew show good agreement, suggesting that \IMRTHM is
accurate enough for GW190814-like events at current SNRs. The constraints on
the component masses and $\chieff$ improve for \IMRTPHM compared to the
nonprecessing models, suggesting that precession should be included in
\NRSurNew.
}
\label{fig:cornerplots}
\end{figure*}

\section{Reanalyzing GW190814}
\label{sec:reanalyzing_GW190814}

\NRSurNew is targeted towards GW events like GW190814~\cite{Abbott:2020khf},
with mass ratios $q\gtrsim9$. As \NRSurNew is more accurate than alternative
models in this region, we now reanalyze GW190814 with \NRSurNew. In addition,
we consider two phenomenological models, \IMRTHM~\cite{Estelles:2020twz} and
\IMRTPHM~\cite{Estelles:2021gvs}. Both of these models include the effects of
subdominant modes, but only \IMRTPHM includes precession effects. Precession
effects are included in \IMRTPHM by ``twisting'' the frame of the nonprecessing
model \IMRTHM to mimic orbital precession~\cite{Estelles:2021gvs}. The GW190814
discovery paper~\cite{Abbott:2020khf} instead considered the
\EOBPHM~\cite{Ossokine:2020kjp} and \IMRPHMOld~\cite{Khan:2019kot} binary BH
models, both of which include the effects of subdominant modes and precession
(through a similar twisting procedure). For simplicity, we do not consider
these models here, but we have verified that our results with \IMRTPHM are
consistent with Ref.~\cite{Abbott:2020khf}. Ref.~\cite{Abbott:2020khf} also
considered models~\cite{Matas:2020wab, Thompson:2020nei} with tidal effects,
but found no measurable tidal signatures; therefore, we only show results for
binary BH models.

Source properties can be inferred from GW data following Bayes' theorem (see
e.g.~Ref.~\cite{Thrane:2019pe} for a review). We analyze the GW190814 data made
public by the LIGO-Virgo-Kagra Collaboration~\cite{Abbott:2020khf,
GW_open_science_center}, using the \texttt{Parallel Bilby}~\cite{Smith:2019ucc}
parameter estimation package with the \texttt{dynesty}~\cite{dynesty_paper}
sampler. Following Ref.~\cite{LIGOScientific:2021djp}, we choose a prior that
is uniform in detector frame component masses, and isotropic in sky location
and binary orientation. For the distance prior, we use the
\texttt{UniformSourceFrame} prior~\cite{Romero-Shaw:2020owr} assuming a
cosmology from~\cite{Planck:2015fie} as implemented in
\texttt{Astropy}~\cite{Astropy:2013muo, Price-Whelan:2018hus}.

When using the nonprecessing models \NRSurNew and \IMRTHM, we use the
\texttt{AlignedSpin} prior~\cite{Romero-Shaw:2020owr, Callister:2021gxf}, with
$-0.5 \leq \chi_{1z}\leq 0.5$ and $\chi_{2z}=0$.  The \texttt{AlignedSpin}
prior follows the generic-spin assumptions of a prior that is uniform in
magnitude and isotropic in orientation for each of the two spin vectors, which
in the nonprecessing case is projected onto the orbital angular momentum. Even
though \IMRTHM allows generic aligned-spins on both BHs, we restrict the model
to the same spin range as \NRSurNew for easy comparison.
We have, however,
verified that using unrestricted aligned-spins for \IMRTHM has a negligible
impact on GW190814 posteriors; this is expected as Ref.~\cite{Abbott:2020khf}
placed a constraint of $\chi_1\lesssim 0.07$ at 90\% credibility, and found
that $\chi_2$ cannot be constrained for GW190814.
When using the precessing
model \IMRTPHM, our prior is uniform in spin magnitudes (with $0 \leq
\chi_{1},\chi_{2}\leq 1$) and isotropic in spin orientations for both BHs. The
reason for considering a precessing model with no spin restrictions is to gauge
the impact of neglecting precession in \NRSurNew.

Figure~\ref{fig:cornerplots} shows posterior distributions for the GW190814
source parameters obtained using \NRSurNew, \IMRTHM and \IMRTPHM. We show
constraints on the source-frame component masses $\mSrc{1}$ and $\mSrc{2}$, the
effective spin $\chieff$, the source-frame chirp mass $\MchirpSrc = \MSrc \,
\eta^{3/5}$, the luminosity distance $D_L$, and cosine of the inclination angle
$\theta_{JN}$ between the total angular momentum $\bm{J}$ and the line of sight
direction $\hat{\bm{N}}$. As \NRSurNew is significantly more accurate (see
Fig.~\ref{fig:ligo_mis}), the differences between \NRSurNew and \IMRTHM can be
used to gauge systematic uncertainties in \IMRTHM. In
Fig.~\ref{fig:cornerplots} we find good agreement between \NRSurNew and
\IMRTHM for all parameters shown, which suggests that semi-analytical models
like \IMRTHM are accurate enough for events like GW190814. However, this may
not be the case as detector sensitivity improves and GW190814-like signals are
observed at larger SNRs. At larger SNRs, the differences noted in
Figs.~\ref{fig:ligo_mis} and \ref{fig:error_histgoram} can become significant.

Finally, comparing the posteriors for \IMRTHM and \IMRTPHM in
Fig.~\ref{fig:cornerplots}, we find that including the effects of precession
leads to stronger constraints on the component masses and $\chieff$, while the
chirp mass, distance and inclination constraints are not significantly
affected. This is in agreement with Ref.~\cite{Abbott:2020khf}, and implies
that precession effects should be included in \NRSurNew. While this can be done
by a frame twisting procedure similar to \IMRTPHM, this method does not capture
the full effects of precession like the asymmetries between pairs of $(\ell,
m)$ and $(\ell, -m)$ spin-weighted spherical harmonic
modes~\cite{Varma:2019csw, Varma:2021csh}. While precessing NR surrogate
models~\cite{Varma:2019csw} capture these effects, they require $\gtrsim 1000$
NR simulations, which are not currently possible at large mass ratios.
Therefore, we leave this exploration to future work.

\section{Conclusion}
\label{sec:conc}
We present \NRSurNew, a surrogate waveform model targeted at large mass ratio
GW events like GW190814. The model is trained on 51 binary BH hybrid waveforms
with mass ratios $q\leq 15$ and aligned spins $ \chi_{1z} \in [-0.5, 0.5]$,
$\chi_{2z}=0$, includes the (2,2), (2,1), (3,3), (4,4), and (5,5) spin-weighted
spherical harmonic modes, and spans the entire LIGO-Virgo bandwidth (with $\fLow=20$
Hz) for total masses $M \gtrsim 9.5 \, M_{\odot}$.  Through a leave-one-out
study, we show that \NRSurNew accurately reproduces the hybrid waveforms, with
mismatches below $\sim 2 \times 10^{-3}$ for total masses $10 \, M_{\odot} \leq
M \leq 300 \, M_{\odot}$. This is at least an order-of-magnitude improvement
over existing semi-analytical models.  The model is made publicly available
through the easy-to-use Python package \textit{gwsurrogate}~\cite{gwsurrogate}.

We reanalyze GW190814 using \NRSurNew and find results consistent with the
discovery paper Ref.~\cite{Abbott:2020khf}. This suggests that current
semi-analytical models are accurate enough for events like GW190814. However,
as detector sensitivity improves, we can expect to see similar signals at a
higher SNR. We anticipate that accurate models like \NRSurNew will be necessary
for analyzing such signals. With that goal, we identify precession as an
important feature to be added to \NRSurNew in the future.

\begin{acknowledgments}
We thank Hector Estelles and Alessandro Nagar for comments on the manuscript.
This work was supported in part by the Sherman Fairchild Foundation and by
National Science Foundation (NSF) Grant Nos. PHY-2011961, PHY-2011968, and
OAC-1931266 at Caltech, and NSF Grant Nos. PHY-1912081 and OAC-1931280 at
Cornell.
V.V. acknowledges funding from the European Union’s Horizon 2020 research and
innovation program under the Marie Skłodowska-Curie grant agreement No.~896869.
V.V. was supported by a Klarman Fellowship at Cornell.
C.-J.H. acknowledges support of the NSF and the LIGO Laboratory.
NR simulations were conducted on the Frontera computing project at the Texas
Advanced Computing Center.  Additional computations were performed on the
Wheeler cluster at Caltech, which is supported by the Sherman Fairchild
Foundation and by Caltech; and the High Performance Cluster at Caltech.
This material is based upon work supported by NSF's LIGO Laboratory which is a
major facility fully funded by the NSF.
This research made use of data, software and/or web tools obtained from the
Gravitational Wave Open Science Center~\cite{GW_open_science_center}, a service
of the LIGO Laboratory, the LIGO Scientific Collaboration and the Virgo
Collaboration.

\end{acknowledgments}

\bibliography{References}

\end{document}